\documentclass[aps,prb,preprint,12pt]{revtex4-1}  %Required for AJP (divide by 3 for # AJP 2-col pages)
\usepackage{amsmath}  % needed for \tfrac, \bmatrix, etc.
\usepackage{amsfonts} % needed for bold Greek, Fraktur, and blackboard bold
\usepackage{graphicx} % needed for figures
\usepackage{color}

\newcommand{\be}[1]{\begin{equation}\label{#1}}
\newcommand{\ee}{\end{equation}}
\newcommand{\bea}[1]{\begin{eqnarray}\label{#1}}
\newcommand{\eea}{\end{eqnarray}}
\newcommand{\no}{\nonumber \\}
\newcommand{\Fig}[1]{Fig.(\ref{#1})}

\newcommand{\Eq}[1]{Eq.(\ref{#1})}

\newcommand{\Sec}[1]{Section~\ref{#1}}
\newcommand{\bsub}{\begin{subequations}}
\newcommand{\esub}{\end{subequations}}
\newcommand{\bwt}{\begin{widetext}}
\newcommand{\ewt}{\end{widetext}}

\usepackage{color}

\def\trm#1{\textrm{#1}}
\def\tit#1{\textit{#1}}

\newcommand{\om}{\omega}
\newcommand{\Om}{\Omega}

\def\a0{{\alpha_0}}

\def\da0{{\dot{\alpha}_0}}

\def\myoverDefn#1#2{\hbox{\space \raise-2mm\hbox{$\textstyle{#1} \atop \scriptstyle{#2}$} }}
\def\defn{\overset{\textrm{def}}{=}}
\def\om{{\omega}}
\def\k{{\kappa}}
\def\t{{\tau}}

\def\mA{{\mathcal{A}}}

\def\mF{{\mathcal{F}}}
\def\mO{{\mathcal{O}}}

\def\G{{\Gamma}}

\def\D{{\Delta}}
\def\Dt{{\Delta t}}
\def\g{{\gamma}}
\def\d{{\delta}}
\def\a{{\alpha}}

\def\Im{\textrm{Im}}

\def\rp{r_{P}}
\def\rp2{r_{p}^{2}}

\newcommand{\half}{\frac{1}{2}}

\newcommand{\pd}{\partial}

\def\eps{\epsilon}
\def\a0{{a_0}}
\def\b_0{{\beta_0}}
\def\b{{\beta}}

\def\etal{\tit{et al.}}
\begin{document}
%====================================
% Be sure to use the \title, \author, \affiliation, and \abstract macros
% to format your title page.  Don't use lower-level macros to  manually
% adjust the fonts and centering.

\title{WKB-like approach to the Unruh temperature\\ for arbitrary acceleration}
% In a long title you can use \\ to force a line break at a certain location.

%When submitting the manuscript for review, do not include the author's name or institution
\author{Paul M. Alsing}
\email{alsingpm@gmail.com, palsing@fau.edu} % optional
%\altaffiliation[permanent address: ]{101 Main Street, Anytown, USA} 
% optional second address
% If there were a second author at the same address, we would put another 
% \author{} statement here.  Don't combine multiple authors in a single
% \author statement.
\affiliation{Department of Physics, Florida Atlantic University, Boca Raton, FL, 33431 }
% Please provide a full mailing address here.

%\author{David P. Jackson}
%\email{ajp@dickinson.edu}
%\affiliation{Department of Physics, Dickinson College, Carlisle, PA 17013}

% See the REVTeX documentation for more examples of author and affiliation lists.

\date{\today}

\begin{abstract}
In this work we study the Unruh temperature as arising from tunneling through a barrier for an observer in flat Minkowski spacetime with arbitrary acceleration $a(t)$. For the defining case of constant acceleration $a(t) = a_0$, the Unruh temperature \cite{Unruh:1976} is given by 
$k_b\,T_U =\tfrac{\hbar\,a_0}{2\,\pi\,c}$. Extending the work of de Gill \tit{et al.} \cite{deGill:2010}, we generalize the gravitational WKB approach to derive the Unruh temperature for arbitrary acceleration.
We show that the often employed Schwarzschild-like form of the flat metric is not appropriate for the WKB calculation with an arbitrary $a(t)$, and instead derive a generalized Unruh temperature for the generalized Rindler metric where $a_0\to a(t)$. We derive a generalization of the Rindler coordinates appropriate for arbitrary $a(t)$, and stress the importance of the role of the integrated acceleration $\chi(t) \defn \int^t dt'\,a(t')$, which can also act as a temporal coordinate. We explore several non-trivial examples of $a(t)$ and their generalized Unruh temperatures. We additionally develop an approximation to the Unruh temperature for small deviations away from constant acceleration by the standard approach of considering the negative frequency content of a purely positive frequency plane wave of an inertial observer, as measured by the co-moving arbitrarily accelerated observer. Lastly, we develop and explicit coordinate transformation between the arbitrarily accelerated observer and conformal coordinates, where the plane wave structure of the solutions of the wave equation is readily transparent, and analogous to the form for the inertial observer.
\end{abstract}
% AJP requires an abstract for all regular article submissions.
% Abstracts are optional for submissions to the "Notes and Discussions" section.

\maketitle % title page is now complete

%=========================================================
\section{Introduction}\label{sec:Intro}
%=========================================================
The Unruh temperature\cite{Unruh:1976} is the effect that a observer undergoing constant acceleration $a_0$ in flat Minkowski spacetime will perceive a thermal flux of particles at a temperature 
$T_U = \tfrac{\hbar\,a_0}{2\,\pi\,c}$, while an inertial observer with zero acceleration will not.  Unruh's 1976 famous effect was published one year after Hawking's celebrated result\cite{Hawking:1975} that black holes (BHs) radiate a thermal flux of particles for a stationary observer sitting just outside, at fixed radial coordinate, from an eternal BH. These effects are intimately related, since in the gravitational case the stationary observer is undergoing constant acceleration $\k = G\,M/R^2_s$ due to the surface gravity of its horizon at the Schwarzschild radius $R_s = 2\,G\,M/c^2$. Further, expanding the Schwarzschild metric close to $R_s$ brings the radial and temporal portions into the Rindler constant acceleration form, and thus the Hawking temperature is given by the Unruh temperature with the replacement $a_0\to\k$.

The Unruh effect has been extensively studied over the years (see Birrell and Davies\cite{Birrell_Davies:1982}, Unruh and Wald\cite{Unruh_Wald:1982}, 
and especially the extensive review article by Crispino \etal\cite{Crispino:2008} and reference therein). The Unruh temperature is most commonly derived by considering a massive or massless scalar field and writing down the expansion of the quantum field in terms of a complete orthonormal set of the positive and negative frequency plane wave solutions (modes) of the classical wave equation associated with the metric, for both the inertial observer Bob, and for the Rindler constant accelerated observer, Rob. While the field is the same regardless of the coordinates, the plane wave solutions for Bob and Rob are not, so that one set of modes can be expanded in terms of the other.
This leads to a Bogoliubov transformation that relates the Bob's annihilation operator to a linear combination of both of Rob's annihilation and creation operator, and visa versa for Rob. Thus, Rob's detectors will measure a non-zero expectation value for his number operator as he moves through the inertial (Minkowski) vacuum of Bob.
Bob, of course, will register a zero expectation value for his number operator for his vaccuum state.
In brief, Rob will measure (via a Fourier transform) negative frequency components for a purely positive frequency plane wave of the inertial Bob, stemming from the fact that Rob's proper time differs Bob's\cite{Alsing_Milonni:2004, Padmanabhan:2010,Gelis:2021}.
Ultimately, this situations arises since for constant acceleration, Rob's orbit is solely confined to the \tit{Right Rindler Wedge} (RRW) $X>|T|>0$ in Bob's inertial coordinates $(T,X)$ with metric $ds^2 = -dT^2 + dX^2$.
These orbits are causally disconnect from any ``mirror" constant accelerated observer traveling in the 
Left Rindler Wedge (LRW), defined by $-|T|< X<0$. Quantum mechanically, the Minkowski vacuum of Bob  is a pure two-mode squeezed state\cite{Scully_Zubairy:1997,Agarwal:2013,Gerry_Knight:2023},  where the ``signal/idler" modes of the squeezed state vaccum represent the  equally populated (Fock) number states of the RRW and LRW, respectively. Tracing out the LRW modes (since no signal from the LRW can ever reach the RRW), leaves a thermal state density matrix with the Unruh temperature. The interpretation carries over immediately to the case of the Hawking effect (see Chapter 9 of Carroll\cite{Carroll:2004} where the RRW and LRW take on the interpretation of the exterior regions outside the horizon, of the extended Penrose spacetime diagram for an eternal BH).
Thus, operationally, without all the quantum field theory machinery, once can understand, and calculate, the Unruh (and Hawking) temperature by simply Fourier analyzing, with respect to the accelerated observers proper time, the frequency content of a purely positive frequency inertial plane wave (in fact, this is the heart of Hawking's original calculation\cite{Hawking:1975}). 

At the beginning of the this century, an alternative approach for computing the Unruh and Hawking temperature has gained popularity by its ease of computation and wide versatility. This method views the Unruh/Hawking temperature as described by a tunneling process arising from complex paths around a pole (for a trajectory going from outside to inside the horizon) in the action $S$ when the scalar field is described as $\phi\sim e^{i S/\hbar}$. The pole is at the singularity of the metric, which indicates the location of the horizon (particle horizon for Unruh, apparent horizon for Hawking). This method was put forth by Srinivasan and Padmanabhan and coauthors\cite{Srinivasan:1999,Srinivasan:2001}, 
and Parikh and Wilczek\cite{Wilczek:2000}, (for an extensive review see Vanzo \etal\cite{Vanzo:2011} and references therein, and recent work related to the BH information loss problem by Zhang \etal\cite{Corda:2025}). The popularity of this method also stems from the ease in which one can incorporate particles of various spin (see Kerner and Mann\cite{Kerner_Mann:2008}, and the informative review monograph by Mann\cite{Mann:2015} with scalar and spin $1/2$ examples).

Particularly relevant to this work is that of de Gill \etal\cite{deGill:2010} in which the authors point out there are two contributions to the expression for the temperature; one from the spatial contribution of the action as the contour goes around the pole, and another  temporal contribution arising from the interchange of temporal/spatial coordinates to spatial/temporal, respectively, as one travels from outside to inside the horizon. The work of de Gill \etal\, considered the Rindler case of constant acceleration $a_0$ in flat (Minkowski) spacetime. In this work we generalize their results to the case of an arbitrary accelerated observer $a(t)$ in flat spacetime. We develop a generalized Unruh temperature $T_U|_{a_0\to a(t)}$ modified by a denominator that involves an integration over both the accelerated profile $a(t)$ and the integral of the acceleration profile $\chi(t)\defn \int dt\,a(t)$.
(Interesting, recent work related to this present work, but not using a WKB approach, has been conducted by 
Hari and Kothawala\cite{Hari_Dawood:2021} and by Hari, Barman and Kothawala\cite{Hari_Barman_Dawood:2025}, and is discussed more fully in the Conclusion).

The outline of this paper is as follows:
in \Sec{sec:Rindler metric} we review the Rindler metric and coordinates for constant acceleration, and generalize them to the case of arbitrary acceleration.
In \Sec{sec:WKB} we review the WKB-approach to the Unruh temperature, briefly reviewing the results of de Gill \etal \cite{deGill:2010}. We then extend the integral expression for the action to arbitrary acceleration and produce a formula generalizing the Unruh temperature for constant acceleration (which reduces to it in the limit $a(t)\to a_0$).
In \Sec{sec:Examples} we consider two acceleration profiles, the first which contains a ``burst" of acceleration at the origin of time, with the acceleration trailing off to zero at large absolute times, and second, an acceleration with a hyperbolic tangent profile so that it is essentially constant $\pm a_0$ for large  (negative/positive) times, with a rapid transition from negative to positive at the temporal origin. 
In \Sec{sec:Unruh:Integral} we examine the Unruh temperature by the conventional approach by Fourier transforming a purely positive frequency inertial plane wave (Bob) with respect to the accelerated observer's proper time (Rob). This yields the positive/negative frequency content of the Bob's plane wave as measured by Rob. We examine the acceleration profiles examined in  \Sec{sec:Examples}.
Lastly, in \Sec{sec:Conformal:Coords} we develop the explicit coordinate transformation between the accelerated observer and conformal coordinates where the metric now has the form of the inertial flat spacetime (Bob) 
$ds^2 = -dT^2 + dX^2$, multiplied by an overall function of time and space. With this conformally flat metric the plane wave structure of the solutions of the wave equation for the accelerated metric is made readily apparent, and analogous to the form for the inertial observer. 
We end in \Sec{sec:Conclusion} with our Conclusion, discussion of recent related work, and prospects for extending this work to curved spacetime.

\section{The Rindler metric for arbitrary acceleration}\label{sec:Rindler metric}
%=========================================================
The standard form of the Rindler metric\cite{Rindler:1956} with coordinates $(t,x)$ (Rob) is given by
\be{Rindler:metric:a0}
ds_R^2 = -(1 + a_0\,x)^2\,dt^2 + dx^2
\ee
which is related to the flat Minkowski metric with inertial coordinates $(T,X)$ (Bob) with
$ds_M^2~=~-dT^2~+~dX^2$ by the Rindler transformation
\be{Rindler:coords:t:x}
T = \frac{1+a_0\,x}{a_0}\,\sinh(a_0\,t), \qquad X = \frac{1+a_0\,x}{a_0}\,\cosh(a_0\,t).
\ee
The co-moving frame of the accelerated observer (Rob) is taken to be $x=0$.
One can always transform the above to a conformally flat form by defining 
 another spatial coordinate $\xi$ by $dx = (1+a_0\,x)\,d\xi$
or $(1+a_0\,x) = e^{a_0\,\xi}$, which allows us to put the Rindler transformation into a modified form
of \Eq{Rindler:coords:t:x}, with the conformally flat metric
\be{Rindler:metric:conformally:flat}
ds_R^2 = e^{2 a_0 \xi}\,(-dt^2 + d\xi^2), \quad 
T = \frac{1}{a_0}\,e^{a_0 \xi}\, \sinh(a_0 t), \quad 
X = \frac{1}{a_0}\,e^{a_0 \xi}\, \cosh(a_0 t).
\ee
This shows that the Rindler metric admits (unnormalized) positive frequency $e^{-i\,\Om\,t}$ and 
negative frequency $e^{i\,\Om\,t}$ plane waves ($\Om>0$), analogous to the positive and negative frequency inertial plane waves given by $e^{\pm i\,\om\,U}$, where $U \defn T-X$. 
In the standard derivation \cite{Unruh:1976,Alsing_Milonni:2004,Padmanabhan:2010}, the Unruh temperature arises from the consideration of the spectrum of the negative frequency content $e^{i\,\Om\,t}$, as seen by Rob, of the purely positive frequency plane wave $e^{- i\,\om\,U}$ of the inertial Bob, which is found to be proportional to the Planck spectrum 
$\tfrac{1}{e^{2\pi\Om/(a_0/c)}-1} \equiv \tfrac{1}{e^{\hbar\,\Om/(k_b\,T_U)}-1}$ with 
$k_b\,T_U \defn \tfrac{\hbar\,a_0}{2\,\pi\,c}$. In the quantum realm, the Unruh temperature arises from the inequivalence of the Rindler vacuum from the standard Minkowski vacuum, so that Rob observes particles within the Minkowski vacuum\cite{Unruh:1976,Birrell_Davies:1982,Carroll:2004} while, of course, the inertial Bob does not.

%===========================================================================
\subsection{Coordinates for an observer in flat spacetime with arbitrary acceleration $\mathbf{a(t)}$}\label{subsec:Rob:coords}
%===========================================================================
Let now generalize the above  standard Rindler coordinates to arbitrary acceleration $a_0~\to~a(t)$.
To achieve this, we begin with the form of inertial Bob's $(T,X)$ 4-velocity $u^\mu(t)$ description of the arbitrary accelerated $a(t)$ Rob $(t,x)$, in terms of his proper time $t$ viz 
\bea{Tprime:Xprime:t}
u^\mu(t) = \big(\dot{T}(t), \dot{X}(t) \big)&\equiv&
\left(\frac{dT(t)}{dt}, \frac{dX(t)}{dt}\right) \defn \big(\g(t), \g(t)\,\b(t)\big) = \big(\cosh\chi(t), \sinh\chi(t)\big), \label{Tprime:Xprime:t:line1} \\
\trm{where}\qquad \g(t)&\defn&\frac{1}{\sqrt{1-\b^2(t)}}, \qquad 
\chi(t)\defn \int^t dt\, a(t) \;\Rightarrow\; \dot{\chi}(t) = a(t). \label{Tprime:Xprime:t:line2} 
\eea
The first form of \Eq{Tprime:Xprime:t:line1}  is chosen to identically satisfy the the unit magnitude condition fo the 4-velocity $u^\mu(t)\, u_\mu(t) = \eta_{\mu\nu}u^\mu(t)\,\,u^\nu(t) = -1$ under the Minkowski metric
$\eta_{\mu\nu} = \trm{diagonal}(-1,1)$, using $\g^2(t) - \b^2(t)\,\g^2(t)\equiv 1$.
The second form of \Eq{Tprime:Xprime:t:line1} was chosen in order to give the proper acceleration 
$a(t) \defn \sqrt{a^\mu(t)\, a_\mu(t)} = \sqrt{\dot{u}^\mu(t)\, \dot{u}_\mu(t)}$, using the second part of 
\Eq{Tprime:Xprime:t:line2}, $ \dot{\chi}(t) = a(t)$, viz
\be{accel}
a^\mu(t) = \dot{u}^\mu(t) = \big(\ddot{T}(t), \ddot{X}(t) \big) =  a(t)\,\big(\sinh\chi(t), \cosh\chi(t)\big).
\ee
Further, from the definition in \Eq{Tprime:Xprime:t:line1} of $\g(t) = (1-\b^2(t))^{-1/2}\defn\cosh\chi(t)$ we can solve for the velocity profile ($c=1$) $\b(t)$ to find
\bea{beta:t}
-1\le \b(t) &=& \tanh(\chi(t)) \le 1, \qquad \chi(t)\defn \int^t dt\, a(t), \label{beta:t:line1} \\
\trm{equivalently}\;\;\Rightarrow a(t) &=& \g^2(t) \, \dot{\b}(t). \label{beta:t:line2} 
\eea 
Thus, for an arbitrary acceleration profile $a(t)$, Rob's velocity profile $\b(t) = v(t)/c$ remains less than $1$ in magnitude, respecting special relativity. By differentiating $\b(t)$ in \Eq{beta:t:line1} and using
$\dot{\chi}(t) = a(t)$ we can write $a(t)$ in \Eq{beta:t:line2} as a function of $\b(t)$ and $\dot{\b}(t)$. 
This shows that we could also ``go the other way around," namely we can equivalently specify an arbitrary velocity profile $-1\le\b(t)\le 1$, and hence determine the corresponding proper acceleration $a(t)$. In the author's recent work \cite{Alsing_I2R:2026} we considered an observer with 
beginning and ending velocity of $\beta_0<1$, with the profile $\b(t) = \b_0\,\tanh(a_0\,t)$. This produced 
a two parameter $(\b_0, a_0)$ generalization of the Rindler coordinates that smoothly interpolated 
between constant acceleration for $\b_0\to 1$, and inertial coordinates for $\b_0\to 0$.
The proper acceleration was given by 
$a(t) = a_0\, \b_0\, \frac{\g_0^2}{\g_0^2+ \sinh^2(a_0\,t)}$.
This current work further extends the previous study to the case of arbitrary acceleration.

For the Rindler case with $a(t)= a_0$ we have $\chi(t) = a_0\,t$ and thus $\b(t) = \tanh(a_0 t)$ which follows from the Rindler transformation \Eq{Rindler:coords:t:x} for Rob, 
$\b(t) = \left.\tfrac{\dot{X}(t)}{\dot{T}(t)}\right|_{x=0}$.
Note that here,  for $t\in\{-\infty, \infty\} \Rightarrow \chi\in\{-\infty, \infty\} \Rightarrow \b(t)\in[-1,1]$, i.e. the constant accelerated Rob begins and ends with a velocity at the speed of light.

 For general acceleration $a(t)$ this need not be the case, i.e. in general 
$t\in\{-\infty, \infty\} \Rightarrow \chi\in\{-\chi_{max} \chi_{max}\}\Rightarrow \b\in[-\tanh(\chi_{max}),\tanh(\chi_{max})]\defn [-\b_{max},\b_{max}]\subset [-1,1]$. That is, Rob with acceleration $a(t)$, will in general, begin and end with a velocity less than the speed of light, $|\b_{max}| <1$. 

As a specific example (that we will further examine later), consider the motion of Rob that begins and ends with zero acceleration, and peaks in the middle at $t=0$ at $a_0$ given by
\bea{at:inv:coshsqrd}
a(t) &=& \frac{a_0}{\cosh^2(b_0\,t)} \overset{b_0\to 0}{\longrightarrow} a_0 \qquad\Rightarrow\quad
\chi(t) = \frac{a_0}{b_0}\, \tanh(b_0\,t) \overset{b_0\to 0}{\longrightarrow} a_0\,t, \label{at:inv:coshsqrd:line1}
\\
\Rightarrow
\b(t) &=& \tanh\Big(\frac{a_0}{b_0}\, \tanh(b_0\,t) \Big) \overset{b_0\to 0}{\longrightarrow} \tanh(a_0\,t), 
\quad \b_{max} = \tanh(a_0/b_0)<1\overset{b_0\to 0}{\longrightarrow} 1. \label{at:inv:coshsqrd:line2}
\eea
Thus, in the limit that $b_0\to 0$ we recover the constant acceleration $a_0$ of the standard Rindler observer, but for $b_0>0$ and finite $a_0$, we have $\b_{max} <1$.

%===========================================================================
\subsection{Coordinate transformation between the inertial and arbitrary acceleration frame: 
$\mathbf{(T(t,x), X(t,x))}$}\label{subsec:TX:tx}
%===========================================================================
We can now integrate \Eq{Tprime:Xprime:t:line1} to obtain the Bob coordinates $(T(t), X(t))$ of the accelerated Rob (located at his origin $x=0$) as 
\be{T:X:at:xeq0}
T(t) = \int^t dt\, \cosh\chi(t), \qquad X(t) = \int^t dt\,\sinh\chi(t), 
\ee
from which we see that for constant acceleration $a(t)=a_0\Rightarrow\chi(t) = a_0\,t$ yields the 
standard Rindler coordinates $T = \tfrac{1}{a_0}\sinh(a_0\,t)$ and  $X = \tfrac{1}{a_0}\cosh(a_0\,t)$
from \Eq{Rindler:coords:t:x} with $x=0$.

To find general coordinates $(t,x)$ for the arbitrarily accelerated Rob, we  follow the arguments of 
Padmanabhan \cite{Padmanabhan:2010}, and in particular, Gelis (p341)  \cite{Gelis:2021},
and note that Bob's coordinates $(T,X)$ for  a  general point $(t,x)$ in Rob's frame 
(where Rob's coordinates at $x=0$ are given by $(T(t),X(t))$ in \Eq{T:X:at:xeq0} by Bob)
is given by
\be{Gelis:p341}
\hspace{-0.5in}
\begin{pmatrix}
	T - T(t)  \\
	 X-X(t) \\
\end{pmatrix}_{\trm{Bob}}
=
\begin{pmatrix}
	 \g(t) & \beta(t)\g(t) \\
	\beta(t)\g(t) & \g(t) \\
\end{pmatrix}
\begin{pmatrix}
	0  \\
	 x \\
\end{pmatrix}_{Rob}
= 
\begin{pmatrix}
	 u^0(t) &u^1(t) \\
	u^1(t) & u^0(t) \\
\end{pmatrix}
\begin{pmatrix}
	0  \\
	 x \\
\end{pmatrix}
= 
\begin{pmatrix}
	 u^1 \, x \\
	u^0 \, x \\
\end{pmatrix}
=
\begin{pmatrix}
	 X'(t) \, x \\
	T'(t) \, x \\
\end{pmatrix}.
\ee
Here, $x$ is the end of a rigid rod with one end at Rob's origin, and the other at position $x$,
traveling on  worldlines given by $(T(t), X(t))$ and $(T, X)$, respectively. The Lorentz transformation that takes one from Bob to Rob's instantaneous rest (co-moving) frame, uses the instantaneous velocity 
$\b(t)$, since the velocity is constantly changing at each instant due to the non-uniform acceleration 
$a(t)$.
Therefore, we have
\bea{T:X:at:x}
T(t) &=& \int^t dt\, \cosh\chi(t) + x\,\sinh\chi(t) \equiv \int^t dt\,(1 + a(t)\,x) \cosh\chi(t), \label{T:X:at:x:line1} \\
 X(t) &=& \int^t dt\,\sinh\chi(t) + x\,\cosh\chi(t) \equiv \int^t dt\,(1 + a(t)\,x) \sinh\chi(t), \label{T:X:at:x:line2}
\eea
where in the second equalities in the above we have used
$d\sinh\chi(t) = a(t) \cosh\chi(t)\,dt$ which after integrating gives
$\sinh\chi(t) = \int^t dt\,a(t) \cosh\chi(t)\,dt$, 
and similarly 
$d\cosh\chi(t) = a(t) \sinh\chi(t)\,dt$ which after integrating gives
$\cosh\chi(t) = \int^t dt\,a(t) \sinh\chi(t)\,dt$.
We can now form the metric by forming the total differentials
\bea{Rindler:metric:at}
dT &=& (1 + a(t)\,x)\cosh\chi(t)\,dt + \sinh\chi(t)\,dx, \label{Rindler:metric:at:line1} \\
dX &=& (1 + a(t)\,x)\sinh\chi(t)\,dt + \cosh\chi(t)\,dx,\quad \label{Rindler:metric:at:line2} \\
\Rightarrow
ds_{R,a(t)}^2 &=& -dT^2 + dX^2 = -(1 + a(t)\,x)^2\,dt^2 + dx^2, \label{Rindler:metric:at:line3}
\eea
where the cross terms $dt\,dx$ cancel exactly. Comparing \Eq{Rindler:metric:at:line3}
with the standard Rindler metric for constant acceleration \Eq{Rindler:metric:a0}, we see that 
the only change for arbitrary acceleration $a(t)$ is the substitution $a_0\to a(t)$.

%===========================================================================
\subsection{Metric $\mathbf{ds_{R,a(t)}^2}$ via flat spacetime Riemann tensor consideration}\label{subsec:TX:tx}
%===========================================================================
While the simple end result that  $a_0\to a(t)$ in \Eq{Rindler:metric:at:line3} may appear at first surprising, once can also derive \Eq{Rindler:metric:at:line3} by
assuming a metric of the form $ds^2_{R,f(t,x)}  = -f(t,x)\,dt^2 + dx^2$, and computing the Ricci scalar 
$R=g^{\mu\nu}\,R_{\mu\nu} = (2 f\, f'' - (f')^2)/(2 f) =0$ for a flat spacetime (where $f' = df/dx$), which is satisfied by
$f(t,x) = c_0(t) + c_1(t)\,x + \frac{c_1^2(t)\,x^2}{4\,c_0(t)}$, for arbitrary time-dependent ``constants" of integration $c_0(t)$ and $c_1(t)$.
Upon choosing 
$c_0(t)=1$ and $c_1(t) = 2\,a(t)$, we obtain $f(t,x) = (1 + a(t)\,x)^2$. 
There are only two Riemann tensor components, both proportional to $R$, i.e.
$R^x_{t t x} = R^t_{x t x}= -R/(2\,f) =0$, so they are identically satisfied by $R=0$ for a flat spacetime.
Therefore, the metric $ds_{R,a(t)}^2$ in \Eq{Rindler:metric:at:line3} supports an observer with arbitrary acceleration $a(t)$ in flat spacetime.

For later WKB calculation consideration, it is worthwhile to note that if one were to assume a metric of the Schwarzschild-form $ds_{Schw}^2 \defn -F(\tau,\zeta)\,d\tau^2 + d\zeta^2/F(\tau,\zeta)$, then a similar analysis of Riemann tensor and scalar curvature for flat spacetime reveals that the only possible solution is of the form
$F(\tau,\zeta) = c_0 + c_1\,\zeta$, where now $c_0$ and $c_1$ are constants independent of $\tau$.
In their WKB computation of the Unruh temperature, de Gill \tit{et al.} \cite{deGill:2010} use $F(\tau,\zeta) = 1 + 2\,a_0\,\zeta$, which is related to the accelerated observer's $(t,x)$ coordinates by defining  
$1 + 2\,a_0\,\zeta \equiv (1 + a_0\,x)^2$ to give the constant acceleration Rindler metric $ds_R^2$ in \Eq{Rindler:metric:a0}. Therefore, the Schwarzschild-form of the flat metric \tit{only} supports the observer's motion under constant acceleration $a_0$, and not an arbitrary acceleration $a(t)$.
Thus, throughout this work we will use only  $ds_{R,a(t)}^2$ as given by the more general form of the metric in \Eq{Rindler:metric:at:line3}.

%====================================================================
\section{WKB-type tunneling approach to the Unruh Temperature}\label{sec:WKB}
%====================================================================
In this section we follow previous works that derive both the Unruh and Hawking temperature from a WKB-type approach which purports that the radiation can be understood as a tunneling process for both scalar particles, as well as those with spin (see Kerner and Mann\cite{Kerner_Mann:2008}, Mann\cite{Mann:2015}, and the comprehensive review article by Vanzo \tit{et al.}\cite{Vanzo:2011} and references therein). In particular, we follow and expand upon the presentation by de Gill \tit{et al.}\cite{deGill:2010}, who pointed out an important imaginary time contribution.

The basic idea is the following\cite{deGill:2010,Mann:2015}.
Consider a scalar field in a background metric. These fields represent vacuum fluctuations that permeate the spacetime given by the metric.
In the usual quantum mechanical tunneling process, two separated classical turning points are joined in imaginary time through a classically forbidden region.
 For a black hole (BH), the outgoing particle itself creates the tunneling barrier. It's trajectory is from the inside to the outside of the BH, a classically forbidden process,
 so that the original wave $e^{-i\,\om\,t}$ picks up an imaginary part $t\to t+i\G$, and therefore decays as $e^{-i\,\om\,t -\om\,\G}$.
 
 The simplest way to illustrate this process\cite{Mann:2015} is with a scalar field $\phi$ and to apply the WKB approximation to the Klein-Gordon wave equation, using the ansatz $\phi = \phi_0\,e^{i\,S/\hbar}$, where $S$ is the action, and $\phi_0$ is an irrelevant constant amplitude that does not influence the tunneling rate. The action $S$ is integrated along the classically forbidden trajectory which starts inside the horizon and finishes at the outside observer (typically at infinity). The equation will have a simple pole located at the horizon (particle horizon for Rindler) since the trajectory is classically forbidden. Using standard contour integration techniques, one integrates around the pole, and the complete contour can be made closed since the part of the trajectory that starts outside the BH and continues to the observer will not contribute to the final tunneling probability, and can thus safely be ignored.  The tunneling rate is then given by\cite{Mann:2015}
 \be{Mann:2015:Eq133}
\G \propto |\phi|^2 = e^{-2\,\Im(S)/\hbar} = e^{-2\,\Im(\oint p_x dx)/\hbar} \defn e^{-\beta_{th}\,E},
\ee
for the semi-classical tunneling probability for the emitted particle. The form of the last equality follows from the assumed stationarity of the spacetime, and $\beta_{th} = \tfrac{1}{k_b\, T_{th}}$ is interpreted as the inverse temperature of the BH (Unruh radiation).

de Gill \tit{et al.}\cite{deGill:2010} point out that the the formula for $\G$ in \Eq{Mann:2015:Eq133} only
represents the imaginary contribution arising from the spatial part of the action to obtain the tunneling rate. In addition, there is a temporal contribution $E\,\Im(\Dt_{total})$ arising from the time coordinate as the horizon is crossed (since the temporal and spatial character of the coordinates are interchanged). Thus, the total tunneling rate is modified to
 \be{deGill:11}
\G \propto   e^{-2\,\Im(\oint p_x dx - E\,\Dt_{total})/\hbar} \defn e^{-\beta_{th}\,E}.
\ee
Comparing the arguments of the exponentials in \Eq{deGill:11} we arrive at a formula for the Unruh temperature (setting Boltzmann's constant $k_b=1$)
\be{deGill:13}
T_U = \frac{E\,\hbar}{\Im(\oint p_x dx) - E\,\Im(\Dt_{total})}.
\ee

In the following, we first illustrate this for the Rindler constant acceleration $a_0$ case following Appendix C of de Gill \tit{et al.}\cite{deGill:2010}, and then extend this for the case of arbitrary acceleration $a(t)$.

%====================================================================
\subsection{WKB tunneling rate for the case of constant acceleration $\mathbf{a_0}$}\label{WKB:Rindler:a0}
%====================================================================
The Klein-Gordon equation for a scalar field $\phi$ for an arbitrary (curved) spacetime metric
$ds_{CST}^2 = g_{\mu\nu}(x)\,dx^\mu\,dx^\nu$ (with metric signature $(-,+,+,+)$) is given by\cite{Birrell_Davies:1982,Carroll:2004}
\be{KGEqn:CST}
(\Box-m^2)\phi = 
\left(
\frac{1}{\sqrt{-g}}\pd_\mu\left(\sqrt{-g}\,g^{\mu\nu}\,\pd_\nu \right) - \frac{m^2\,c^2}{\hbar^2}
\right)=0.
\ee
Setting $c=1$ and multiplying by $-\hbar^2$ \Eq{KGEqn:CST} becomes
\be{KGEqn:CST:v2}
-\frac{\hbar^2}{\sqrt{-g}}
\Big[
(\pd_\mu\,\sqrt{-g})\,g^{\mu\nu}\,\pd_\nu\,\phi + \sqrt{-g}\,(\pd_\mu g^{\mu\nu})\,\pd_\nu\,\phi 
+  \sqrt{-g}\, g^{\mu\nu}\,\pd_mu\,\pd_\nu\phi
\Big]
+ m^2\,\phi=0.
\ee
For diagonal metrics, this can be considerably reduced (see Appendix C of de Gill \tit{et al.} \cite{deGill:2010}) by using\cite{DInverno:1992,Ryder:2009}
 (i) the covariant derivative of the metric is zero: 
$\nabla_\alpha g^{\mu\nu} = \pd_\alpha g^{\mu\nu}  + \G^\mu_{\alpha\beta}\,g^{\beta\nu} + \G^\nu_{\alpha\beta}\,g^{\mu\beta} = 0$, 
(ii) $\G^\nu_{\alpha\beta}=0$ for $\mu\ne\alpha$ for diagonal metrics,
(iii) $\G^\mu_{\mu\g} = \pd_\g(\ln\sqrt{-g}) =  \pd_\g(\sqrt{-g})/\sqrt{-g}$,
(iv) which allows us to write $\pd_\mu g^{\mu\nu}$ in \Eq{KGEqn:CST:v2} as
$\pd_\mu g^{\mu\nu} = -(\G^\mu_{\mu\g} g^{\g\nu} + \G^\nu_{\mu\g} g^{\mu\g} ) = -(\pd_\g\sqrt{-g}/\sqrt{-g})\,{g^{\g\nu}}$ when the harmonic condition is imposed on the metric, namely
(v)~$\G^\nu_{\mu\g} g^{\mu\g}=0$.
The net result of all these metric manipulations is to reduce \Eq{KGEqn:CST:v2} to the very simple form
\be{KGEqn:CST:v2}
-\hbar^2 g^{\mu\nu}\pd_\mu\pd_\nu\phi + m^2\phi=0.
\ee
Upon substituting in the ansatz for the scalar field as $\phi =\phi_0\,e^{i\,S/\hbar}$ one obtains
\be{deGill:A8}
-i\,\hbar\,g^{\mu\nu}\pd_\mu\pd_\nu S + g^{\mu\nu}\,(\pd_\mu S)\,(\pd_\nu S) + m^2 = 0.
\ee
Dropping the first term $\sim\mO(\hbar)$, leaves the famous Hamilton-Jacobi\cite{Landau_Lifshitz:CTF:1975} equation of classical physics, or the eikonal equation, which has an optical-mechanical analogy between wavefronts and surfaces of constant action\cite{Alsing_AJP:1998}, with energy-momentum equal to the partial derivative of the action, $p_\alpha(x) = -\pd_\alpha S(x)$.
%, as illustrated in \Fig{fig:S0}(left). 

For the case of the Rindler constant acceleration $a_0$ 
we assume an ansatz for the action as
\be{S:ansatz:a0}
S(t,x) = E\,t - S_0(x) =  E\,t -\int p_x\,dx.
\ee
%=====================================
\begin{figure}[h!] 
\begin{center}
\includegraphics[width=6.0in,height=1.75in]{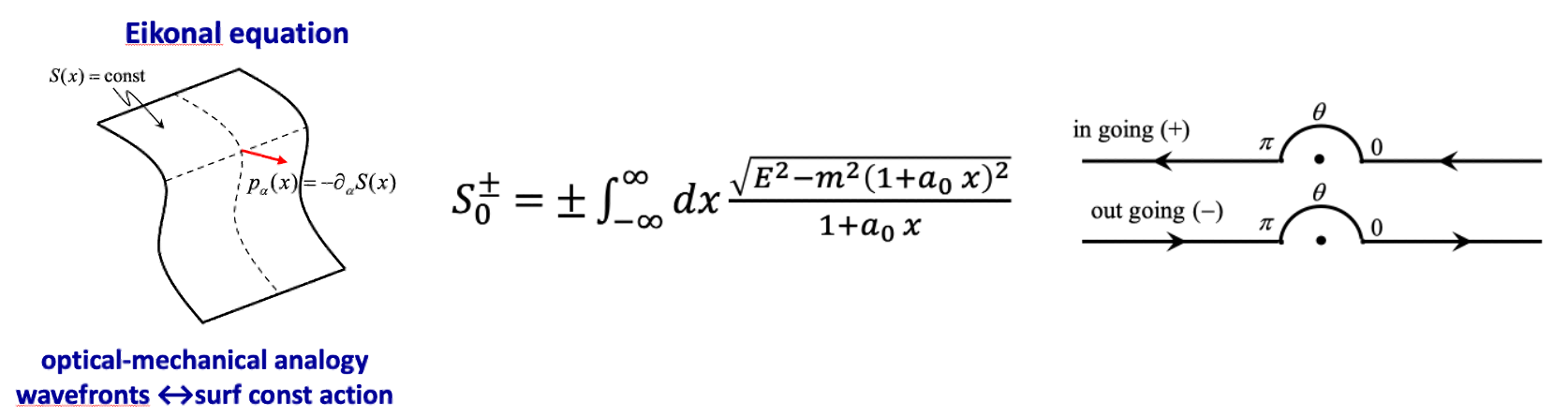} 
\end{center}
\caption{(left) Schematic of surfaces of constant action $S(x)$, 
(middle) integral for $S_0$, and
(right) the contour paths around the pole $x=-1/a_0$ for ingoing and outgoing integration paths.
}\label{fig:S0}
\end{figure}
%=====================================

The Hamilton-Jacobi equation becomes (see \Fig{fig:S0})
\bea{HJ:a0}
0 &=& -\frac{(\pd_t S)^2}{(1+a_0\,x)^2} + (\pd_x S)^2 + m^2 = -\frac{E^2}{(1+a_0\,x)^2} + (\pd_x S_0)^2 + m^2, \label{HJ:a0:line1} \\
&\Rightarrow&  S_0^\pm = \pm\, \int_{-\infty}^{\infty} dx\, \frac{\sqrt{E^2 - m^2\,(1+ a_0\,x)^2}}{1 + a_0\,x}, \qquad \trm{pole at ``horizon"}\; x = -1/a_0. \label{HJ:a0:line2}
\eea
from which we see that there is a pole at the (particle) horizon at $x=-1/a_0$.

To navigate around the pole, we choose $1+a_0\,x = \sqrt{\eps\,e^{i\theta}} =\sqrt{\eps}\,e^{i\theta/2}$, 
$dx = (i/2)\,\sqrt{\eps} e^{i\theta/2}\,d\theta$.
The choice of the square root is explained by de Gill \etal\cite{deGill:2010} as necessary in order for the
Unruh temperature to give the same (standard) result as when using the commonly employed Schwarzschild-form of the Rindler metric $ds^2_{Schw} = -(1+2\,a_0\, \zeta)\,d\tau^2 + (1+2\,a_0\, \zeta)^{-1}\,d\zeta^2$. However, as we have shown in the previous section, the Schwarzschild-form of the metric in flat spacetime \tit{only} supports constant accelerating $a_0$ observers, and so we will use the standard Rindler metric  $ds^2_{R} = -(1+a_0\,x)^2\,dt^2 + dx^2$ for this, and the subsequent arbitrary acceleration $a(t)$ calculation.

Using the above expansion around the pole \Eq{HJ:a0:line2} becomes
\bea{S0:pm:a0}
S_0^+ &=&  \lim_{\eps\to 0}\;  \hspace{0.25in} \int_{0}^{\pi} dx\, 
\frac{\sqrt{E^2 - m^2\,\eps\,e^{i\theta/2}} }{\sqrt{\eps}\,e^{i\theta/2}}\,
(i/2)\,\sqrt{\eps} e^{i\theta/2}\,d\theta = \frac{i\,\pi\,E}{2\,a_0}, \label{S0:pm:a0:line1} \\
S_0^- &=&  \lim_{\eps\to 0}\; -\int_{\pi}^{0} dx\, 
\frac{\sqrt{E^2 - m^2\,\eps\,e^{i\theta/2}} }{\sqrt{\eps}\,e^{i\theta/2}}\,
(i/2)\,\sqrt{\eps} e^{i\theta/2}\,d\theta \hspace{0.1in} = \frac{i\,\pi\,E}{2\,a_0}, \label{S0:pm:a0:line2}
\eea

The temporal contributions to \Eq{deGill:13} arise from\cite{deGill:2010} the fact that 
\bea{slide:15:Taiwan:talk:2011}
x> -1/a_0: &{}&
\begin{array}{c}
T = (a_0^{-1} + x)\, \sinh(a_0\,t) \\
X = (a_0^{-1} + x)\, \cosh(a_0\,t)
\end{array}
\Rightarrow \sinh(a_0\,t)\to \sinh(a_0\,t - i\pi/2) = -i\cosh(a_0\,t), \qquad \label{slide:15:Taiwan:talk:2011:line1} \\
x< -1/a_0: &{}&
\begin{array}{c}
T = (a_0^{-1} + x)\, \cosh(a_0\,t) \\
X = (a_0^{-1} + x)\, \sinh(a_0\,t)
\end{array}
\Rightarrow \cosh(a_0\,t)\to \cosh(a_0\,t - i\pi/2) = -i\sinh(a_0\,t), \qquad \label{slide:15:Taiwan:talk:2011:line2} \\
\Rightarrow &{}& E\,\Dt_{total} = E\,[-i\pi/(2\,a_0) - i\pi/(2\,a_0)] = -i\pi\,E/a_0. \label{slide:15:Taiwan:talk:2011:line3}
\eea
de Gill \etal\ explain the factors of $-i$ as arising from using the form of the Schwarzschild-form of metric
where in  \Eq{slide:15:Taiwan:talk:2011:line1} e.g. 
$a_0^{-1}\,(1 + a_0\,x)\, \sinh(a_0\,t) = a_0^{-1}\sqrt{1+ 2\,a_0\,\zeta}\, \sinh(a_0\,\tau) \to 
 a_0^{-1}\,(i)\,\sqrt{|1+ 2\,a_0\,\zeta|}\, (-i)\,\cosh(a_0\,\tau) = a_0^{-1}\sqrt{|1+ 2\,a_0\,\zeta|}\, \cosh(a_0\,\tau) = a_0^{-1}\,(1 + a_0\,x)\, \cosh(a_0\,t)$, where the first expression 
 is valid for $\zeta>-1/(2\,a_0)$ ($x>-1/a_0)$, 
 and the second expression is valid for $\zeta<-1/(2\,a_0)$ ($x<-1/a_0$) as one crosses the horizon.
 Therefore, we take the expression for $E\,\Dt_{total}$ in \Eq{slide:15:Taiwan:talk:2011:line3} as valid for either form of the Rindler metric.
 Substituting \Eq{S0:pm:a0:line1}, \Eq{S0:pm:a0:line2}, and \Eq{slide:15:Taiwan:talk:2011:line3} into 
 \Eq{deGill:13} one obtains  (de Gill \etal\cite{deGill:2010}) the Unruh temperature (with $k_b, c=1$)
 \be{deGill:21}
T_U = \frac{E\,\hbar}{\Im(S_0^+ + S_0^-) - E\,\Im(\Dt_{total})} 
= \frac{E\,\hbar}{\frac{\pi E}{a_0} - (\frac{-\pi E}{a_0}) }  = \frac{\hbar\,a_0}{2\,\pi}.
\ee

%====================================================================
\subsection{WKB tunneling rate for the case of arbitrary acceleration $\mathbf{a(t)}$}\label{WKB:Rindler:at}
%====================================================================
We now repeat the previous calculation for the  constant acceleration $a_0$ Rindler metric, with the generalized Rindler metric  
$ds_{R,a(t)}^2 = -(1 + a(t)\,x)^2\,dt^2 + dx^2$,
for arbitrary acceleration $a(t)$. 
Because the acceleration is no longer constant, the ansatz for the action in \Eq{S:ansatz:a0} must be changed to
\be{S:ansatz:at}
S(t,x) = E\,t - S_0(t,x). 
\ee
Analogous to \Eq{HJ:a0:line1} and \Eq{HJ:a0:line2}
the Hamilton-Jacobi equation and resulting integral for the action becomes
\bea{HJ:at}
\hspace{-0.5in}
0 &=& -\frac{(\pd_t S)^2}{(1+a(t)\,x)^2} + (\pd_x S)^2 + m^2 = -\frac{(E-\pd_t S_0(t,x))^2}{(1+a(t)\,x)^2} + (\pd_x S_0)^2 + m^2, \label{HJ:at:line1} \\
\hspace{-0.5in}
&\Rightarrow&  S_0^\pm = \pm\, \int_{-\infty}^{\infty} dx\, \frac{\sqrt{\big(E-\pd_t S_0(t,x)\big)^2 - m^2\,(1+ a(t)\,x)^2}}{1 + a(t)\,x}, \qquad \trm{pole at ``horizon"}\; x = -1/a(t).\qquad \label{HJ:at:line2}
\eea
Making the analogous previous substitution, for each fixed $t$, for the integration around the pole, 
now at $x = -1/a(t)$, i.e. 
$1+a(t)\,x = \sqrt{\eps\,e^{i\theta}} =\sqrt{\eps}\,e^{i\theta/2}$, 
$dx = (i/2)\,\sqrt{\eps} e^{i\theta/2}\,d\theta$, 
we have the analogous expression for $S_0^+$ given by
\be{S0:plus:at}
%\hspace{-0.5in}
S_0^+ =  \lim_{\eps\to 0} \int_{0}^{\pi} dx\, 
\frac{\sqrt{\big(E-\pd_t S_0^+(t,x)\big)^2 - m^2\,\eps\,e^{i\theta/2}} }{\sqrt{\eps}\,e^{i\theta/2}}\,
(i/2)\,\sqrt{\eps} e^{i\theta/2}\,d\theta = \frac{i\,\pi\,(E-\pd_t S_0^+)}{2\,a(t)}, 
\ee
and similarly for $S_0^-$.
We treat \Eq{S0:plus:at} as a differential equation for $S_0^+$ for fixed $E$, with the solution
%\be{S0:soln:at}
%\hspace{-0.5in}
%S_0^+ = E \, e^{i\,y(t)} \int^t\, dt'\, e^{i\,y(t')} \overset{a(t)\to a_0}{\longrightarrow}
%E\, e^{i\,2\,a_0\,t/\pi} \, \tfrac{1}{(-i\, 2\,a_0/\pi)} \,e^{-i\,2\,a_0\,t/\pi} = \frac{i\,\pi\,E}{2\,a_0}, \quad
%y(t) \defn \frac{2\,\chi(t)}{\pi}.
%\ee
\be{S0:soln:at}
\hspace{-0.65in}
S_0^+ = E \, e^{i\,y(t)} \int^t\, dt'\, e^{-i\,y(t')}, \quad y(t) \defn \frac{2\,\chi(t)}{\pi}, \quad
S_0^+ \overset{a(t)\to a_0}{\longrightarrow}
E\, e^{i\,2\,a_0\,t/\pi} \, \tfrac{1}{(-i\, 2\,a_0/\pi)} \,e^{-i\,2\,a_0\,t/\pi} = \frac{i\,\pi\,E}{2\,a_0}, 
\ee

For the temporal contribution we have
\bea{slide:19:Taiwan:talk:2011}
x> -1/a(t): &{}&
\begin{array}{c}
T = \int dt (1+ a(t)\,x)\, \cosh\chi(t) \\
X =  \int dt (1+ a(t)\,x)\, \sinh\chi(t)
\end{array}
%\Rightarrow \sinh(a_0\,t)\to \sinh(a_0\,t - i\pi/2) = -i\cosh(a_0\,t), \qquad \\
\\ \label{slide:19:Taiwan:talk:2011:line1}
x< -1/a(t): &{}&
\begin{array}{c}
T = \int dt (1+ a(t)\,x)\, \sinh\chi(t) \\
X =  \int dt (1+ a(t)\,x)\, \cosh\chi(t)
\end{array}
%\Rightarrow \cosh(a_0\,t)\to \cosh(a_0\,t - i\pi/2) = -i\sinh(a_0\,t), \qquad 
\label{slide:19:Taiwan:talk:2011:line2} \\
%
%\Rightarrow &{}& E\,\Dt_{total} = E\,[-i\pi/(2\,a_0) - i\pi/(2\,a_0)] = -i\pi\,E/a_0. \label{slide:15:Taiwan:talk:2011:line3}
\eea
For each fixed instant $t$ we can replace (to lowest order in $\Dt$) $a_0$ by $a(t)$ 
in \Eq{slide:15:Taiwan:talk:2011:line1} and \Eq{slide:15:Taiwan:talk:2011:line2}
to have e.g. $\sinh(a_0 t-i\pi/2) = -i\cosh(a_0 t) \to \sinh\chi(t) = -i\cosh\chi(t)$, that is 
%Thus, to lowest order, we have as in 
%\Eq{slide:15:Taiwan:talk:2011:line1} and \Eq{slide:15:Taiwan:talk:2011:line2}
%that e.g. $\sinh\xi(t)
\be{chi:Dt:slide:19:Taiwan:talk:2011}
\hspace{-0.5in}
\chi(t)\to \chi(t+\Dt) \equiv \chi(t) - i\pi/2  \Rightarrow
\int_t^{t+\Dt}a(t)\, dt  \approx a(t)\Dt = -i\pi/2  \Rightarrow \Dt = -i\pi/(2\,a(t)).
\ee
%\bea{chi:Dt:slide:19:Taiwan:talk:2011}
%\chi(t)&\to& \chi(t+\Dt) = \chi(t) - i\pi/2  dt( a(t) + \dot{a}(t)\Dt)\approx dt\,a(t) \no
%%
%&\Rightarrow&
%\int_t^{t+\Dt}a(t)\, dt  \approx a(t)\Dt = -i\pi/2 \Rightarrow \Dt = -i\pi/(2\,a(t)) 
%\eea
Thus to lowest order in $\Dt$,  we take analogous to \Eq{slide:15:Taiwan:talk:2011:line3}
\be{E:Dt:total:at}
E\,\Dt_{total} = E\,[-i\pi/(2\,a(t)) - i\pi/(2\,a(t))] = -i\pi\,E/a(t). 
\ee
In essence,  we are at each instant $t$  allowing for the same temporal contribution as in the
(locally, instantaneously transformation to a) constant Rindler Schwarzschild-form discussed after \Eq{slide:15:Taiwan:talk:2011:line3}.

Substituting these into \Eq{deGill:21} (with $S_0^-=S_0^+$) we obtain
 \bea{TU:at}
T_U &=& \frac{E\,\hbar}{\Im(S_0^+ + S_0^-) - E\,\Im(\Dt_{total})} 
= 
\frac{ E\,\hbar}
{
\Im\big( 2\,E \, e^{i\,y(t)} \int^t\, dt'\, e^{i\,y(t')}\big) - \big(\frac{-\pi E}{a(t)}\big)
}, \label{TU:at:line1} \\
&=&
\left(\frac{\hbar\,a(t)}{2\,\pi}\right)\,
\frac{1}
{
%\half\big[\Im\big( \, \frac{e^{i\,y(t)}}{a(t)} \int^t\, dt'\, e^{i\,y(t')}\big) +1\big]
\half\big[\Im(\mF) +1\big]
}, \quad \mF \defn \,a(t)\,e^{i\,y(t)} \int^t\, dt'\, e^{-i\,y(t')}
\equiv
\,a(y)\,e^{i\,y} \int^y\, dy' \, \frac{e^{-i\,y'}}{a(y)}. \qquad  \label{TU:at:line2} 
\eea
From the above, we see that Unruh temperature is (i) essentially the same expression as for the case of constant acceleration, but with $a_0\to a(t)$ (the first factor in  \Eq{TU:at:line2}), (ii) modified by a denominator factor $\half\big[\Im(\mF) +1\big]$ that depends on the integrated acceleration.
In the final expression for $\mF$ in \Eq{TU:at:line2}, we have used 
$y = \tfrac{2}{\pi}\,\chi(t) =  \tfrac{2}{\pi}\,\int dt\, a(t)$ 
as the integration variable, and are implying that we can express (i.e. invert) the acceleration  in terms of $y$, i.e. $a(t(y)) = a(y)$. For the Rindler case of constant acceleration $a(t)=a(y)=a_0$ a simple calculation yields $\mF = i$, so that $\Im(\mF)=1$ and the denominator simply becomes unity, reproducing the usual constant acceleration Unruh temperature $T_U = \tfrac{\hbar\,a_0}{2\,\pi}$. In the next section, we explore several examples of \Eq{TU:at:line2} for non-uniform accelerations.

%====================================================================
\section{Examples of Unruh temperature for non-uniform acceleration}\label{sec:Examples}
%====================================================================
In this section, we investigate some examples of the Unruh temperature
\Eq{TU:at:line2} for non-uniform acceleration profiles.

%====================================================================
\subsection{$\mathbf{a(t) = \frac{a_0}{\boldsymbol{\cosh}^2(a_0\,t)} }$}\label{sec:Example:1}
%====================================================================
We return to our example acceleration in \Eq{at:inv:coshsqrd}. However, to keep the relevant integrals 
as analytically tractable as possible, we set $b_0 = a_0$, which no longer allows us to recover the constant acceleration Rindler case $a(t)=a_0$, except in the limit of very small $a_0$.
From \Eq{at:inv:coshsqrd:line1} and \Eq{at:inv:coshsqrd:line2}  we have
\bea{Ex:at:inv:coshsqrd}
\hspace{-0.25in}
a(t) &=& \frac{a_0}{\cosh^2(a_0\,t)} = a_0 \,\big(1-(\pi\,y(t)/2)^2\big)  \qquad\Rightarrow\quad
\chi(t) = \tanh(a_0\,t) \defn \pi\,y(t)/2,  \label{Ex:at:inv:coshsqrd:line1}
\\
\hspace{-0.25in}
\Rightarrow
\b(t) &=& \tanh\big(\tanh(a_0\,t) \big), 
\qquad \b_{max} = \tanh(1)= 0.762. \label{Ex:at:inv:coshsqrd:line2}
\eea
%=====================================
\begin{figure}[h!] 
\begin{center}
%\begin{tabular}{ccc}
%\hspace{-0.5in}
%\includegraphics[width=6.0in,height=2.75in]{fig_TX_orbit_and_T_Unruh_for_a=a0divcoshsqrd(a0t)_21Feb2026} 
\includegraphics[width=6.0in,height=2.75in]{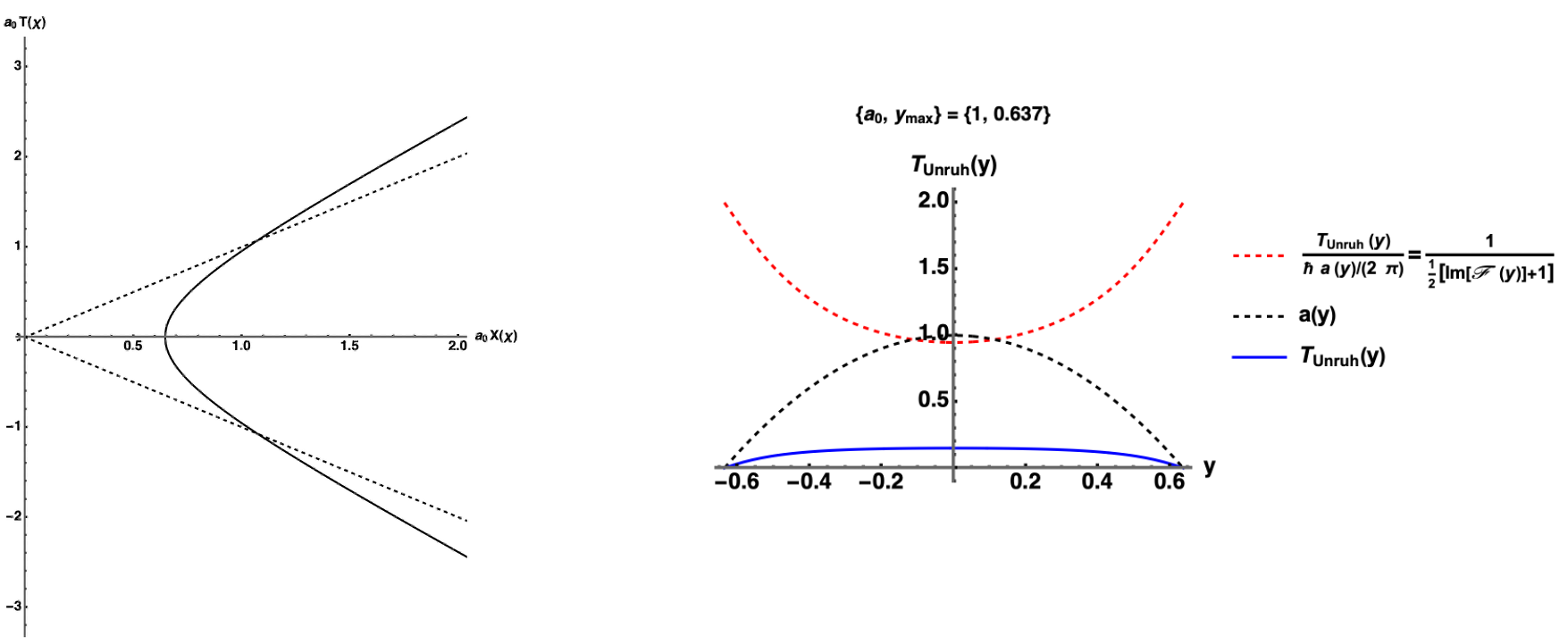} 
%\end{tabular}
\end{center}
\caption{$a(t) = \frac{a_0}{\cosh^2(a_0\,t) }$.
(left): (black) $\big(a_0\,X(t), a_0\,T(t)\big)$ orbit, 
(black, dashed) $T=\pm X$ asymptotes,
(right): (blue) Unruh temperature $T_U(y)$ (\Eq{TU:at:line2}),
 as a function of $y = (2/\pi)\,\chi(t) = (2/\pi)\,\int dt\, a(t)$
(\Eq{Ex:at:inv:coshsqrd:line1}),
for $t\in\{-\infty, \infty\}$.
 (red) Fractional denominator $\frac{1}{\half\Im[\mF(y) +1]}$, and 
 (black, dashed) acceleration $a(y)$ as functions of $y\in[-2/\pi, 2/\pi]$ for  $\chi = \tanh(a_0\,t)\in[-1,1]$.
}\label{fig:a0divcoshSqrda0t}
\end{figure}
%=====================================
The integral $\mF$ in \Eq{TU:at:line2} can be performed and yields
%\be{F:a0divcoshSqrda0t}
%\mF = e^{i y}\,\big(1-(\pi y/2)^2\big)\frac{e^{-i 2/\pi}}{\pi}\,
%\big[
%(-1)\, \trm{Ei}\left(\frac{i(2-\pi y)}{\pi}\right) + 
%e^{i 4/\pi}\, \trm{Ei}\left(\frac{-i(2+\pi y)}{\pi}\right)
%\big]
%\ee
\be{F:a0divcoshSqrda0t}
\mF = \frac{1}{\pi}e^{i y}\,\big(1-(\pi y/2)^2\big)\,
\left[
e^{i\,2/\pi}\, \trm{Ei}\left(\frac{-i(2+\pi y)}{\pi}\right)
-e^{-i\,2/\pi}\,\trm{Ei}\left(\frac{i(2-\pi y)}{\pi}\right)  
\right],
\ee
where Ei$(z) = - \int_{-z}^\infty dt\, t^{-1}\,e^{-t}$ is the exponential integral function, which can be evaluated numerically in software such as \tit{Mathematica}. This was used to evaluate the curves in 
\Fig{fig:a0divcoshSqrda0t}(right). The use of $y$ as a coordinate allows to display this curves in a compact region of $y\in[-2/\pi, 2/\pi]$ for  $\chi = \tanh(a_0\,t)\in[-1,1]$, as $t$ varies over an infinite range 
$t\in[-\infty, \infty]$. We see that the generalized Unruh temperature $T_U(y)$ is essentially constant over all of $y$, but must bend towards zero at the extreme limits $y=\pm 2/\pi$ corresponding to
$a(t) \overset{t\to\pm\infty}{\longrightarrow}=0$.
% $\lim_{t\to\pm\infty} a(t) = 0$

The equations for the orbit for the accelerated observer at $x=0$ can be computed from
$a_0\,T(\chi)=\int\, d\chi\, \cosh(\chi)\, (1-\chi^2)^{-1}$ and 
$a_0\,X(\chi)=\int\, d\chi\, \sinh(\chi)\, (1-\chi^2)^{-1}$ which are parametrically plotted
in \Fig{fig:a0divcoshSqrda0t}(left) for $\chi\in\{-1,1\}$ (black curve), with the lightcone asymptotes
$T = \pm X$ (black, dashed curve). The expressions involve the 
cosh-integral Chi$(z) = \g_E + \log(z) + \int_0^z\,dt\, t^{-1}\,(\cosh(t)-1)$ (where $\g_E$ is Euler's constant),  
and 
sinh-integral Shi$(z)=\int_0^z dt\, t^{-1}\,\sinh(t)$, and are not particularly illuminating, yet they can be computed numerically in \tit{Mathematica}.

%====================================================================
\subsection{$\mathbf{a(t) =a_0\,\boldsymbol{\tanh}(a_0\,t) }$}\label{sec:Example:2}
%====================================================================
As a second example, we consider the acceleration profile 
$a(t) =a_0\,\tanh(b_0\,t)$ that now includes both positive and negative acceleration
$-a_0 \le a(t) \le a_0$.
%%=====================================
%\bea{Ex:a0tanha0t}
%\hspace{-0.5in}
%a(t) &=& a_0\,\tanh(a_0\,t) = \trm{sign}(y)\,a_0 \,\sqrt{1-e^{-\trm{sign}(y) \pi\,y}}  \;\Rightarrow\;
%\chi(t) =  \trm{sign}(y)\,\ln(\cosh(a_0\,t)) \defn \pi\,y(t)/2, \qquad \label{Ex:a0tanha0t:line1}
%\\
%%
%\hspace{-0.5in}
%\Rightarrow
%\b(t) &=& \frac{\sinh^2(a_0\,t)}{\cos^2(a_0\,t)+1}, 
%\qquad \b_{max} = 1. \label{Ex:a0tanha0t:line2}
%\eea
%%=====================================
\bea{Ex:a0tanha0t}
\hspace{-0.5in}
a(t) &=& a_0\,\tanh(a_0\,t) = \trm{sign}(y)\,a_0 \,\sqrt{1-e^{-\trm{sign}(y) \pi\,y}}, \quad
\trm{sign}(y) = \trm{sign}(t),
\label{Ex:a0tanha0t:line1} \\
\hspace{-0.5in}
\;\Rightarrow\;
\chi(t) &=&  \trm{sign}(t)\,\ln(\cosh(a_0\,t)) \defn \pi\,y(t)/2, 
\quad
\b(t) = \trm{sign}(t)\,\frac{\sinh^2(a_0\,t)}{\cos^2(a_0\,t)+1}, 
\;\; \b_{max} = 1. \label{Ex:a0tanha0t:line2}
\eea
%=====================================
\begin{figure}[h!] 
\begin{center}
%\begin{tabular}{ccc}
%\hspace{-0.5in}
%\includegraphics[width=7.0in,height=2.5in]{fig_TX_orbit_and_T_Unruh_for_a=a0tanh(a0t)_21Feb2026}  
\includegraphics[width=7.0in,height=2.5in]{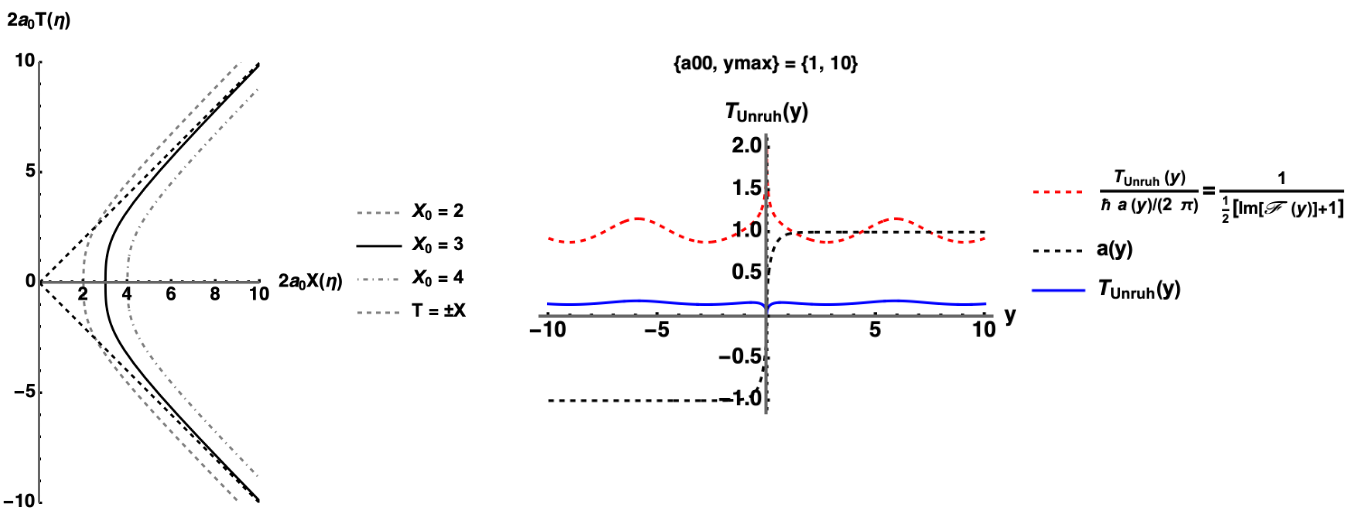}  
%\end{tabular}
\end{center}
\caption{$a(t) = a_0\,\tanh(a_0\,t)$:
(left) (black) $\big(2\,a_0\,X(t), 2\,a_0\,T(t)\big)$ orbit (for $a_0=1$), 
(black, dashed) $T=\pm X$ asymptotes,
(right) (blue) Unruh temperature $T_U(y)$ (\Eq{TU:at:line2}) ,
 as a function of $y = (2/\pi)\,\chi(t) = (2/\pi)\,\int dt\, a(t)$
(\Eq{Ex:at:inv:coshsqrd:line1}),
for $t\in\{-\infty, \infty\}$.
 (red) Fractional denominator $\frac{1}{\half\Im[\mF(y) +1]}$, and 
 (black, dashed) acceleration $a(y)$ as functions of $y\in[-10,10]\subset [-\infty, \infty]$ for  $\chi \in[-\infty,\infty]$.
}\label{fig:a0tanha0t}
\end{figure}
%=====================================

From the expression for the velocity profile $\b(t)$ in \Eq{Ex:a0tanha0t:line2}, we see that 
%$\lim_{t\to\pm\infty}\b(t) = \pm 1$, the speed of light.
$\b(t)\overset{t\to\pm\infty}{\longrightarrow}  \pm 1$, the speed of light.
Therefore, for this acceleration profile the motion is very nearly Rindler-like with 
$a(t)\overset{|t|\gg 1/a_0}{\longrightarrow}  \pm a_0$, which is borne out in the 
$(X,T)$ orbit plots in \Fig{fig:a0tanha0t}(left), where
$2\,a_0\,T(\eta) = \sinh(\eta) + 2\,(\tan^{-1}(e^\eta) - \pi/4)$, and 
$2\,a_0\,(X(\eta)-X_0) = \sinh(\eta) - 2\,(\tan^{-1}(e^\eta) - \pi/4)$,
where $\eta\defn a_0\,t$. The factor of $-\pi/4$ is present to cancel the factor 
$\tan^{-1}(e^\eta)\overset{t\to 0}{\longrightarrow}\pi/4$, to ensure that
$T(\eta=0)=0$ and $X(\eta=0)=X_0$. Thus, we see explicitly that 
$T/X\overset{|t|\to\infty}{\longrightarrow}\pm 1$, asymptoting rapidly to the 
(black, dashed) lightcone in \Fig{fig:a0tanha0t}(left).

Correspondingly in  \Fig{fig:a0tanha0t}(right) we see (black, dashed) acceleration curve as essentially constant at $\pm a_0$ (choosing $a_0=1$), but with a very rapid ``S-curve" transition from $-a_0\to a_0$ around $y=0$ ($t=0$), characteristic of a hyperbolic tangent function. The (blue) generalized Unruh temperature is nearly constant, except for a small cusp-like behavior near the origin (as well as for the denominator factor
$\frac{1}{\half\,[\Im(\mF)+1]}$) due to the brief, rapid transition in the acceleration. The denominator-integral 
$\mF$ can be performed, yielding (for $y>0$)
$\mF(y) = \frac{2\,(e^{\pi\,y}-1)}{\pi - 2\,i}\,{}_2F_1(1, \frac{\pi - i}{\pi}, \frac{3}{2} -  \frac{i}{\pi}, e^{\pi\,y})$,
(with $y\to -y$ for $y<0$),
where ${}_2F_1$ is the hypergeometric function defined by
${}_2F_1(a,b; c;z) = \sum_{k=0}^\infty \frac{(a)_k\,(b)_k}{(c)_k}\,\frac{z^k}{k!}$, where
$(a)_k=a\,(a+~1~)\ldots(a+n-1) = \G(a+n)/\G(a)$ is the Pochhammer symbol. Again, this can be evaluated numerically in \tit{Mathematica}.

Recall that for Rindler case of constant acceleration $a(t)=a_0$ the velocity profile is $\b_R(t) = \tanh(a_0\t)$.
In this example, the velocity profile is nearly $\b_R(t) \approx \trm{sign}(t)\,\tanh^2(a_0\,t)$, for $|t|\gg 1/a_0$,
akin to a ``squared" Rindler velocity profile, and corresponding, nearly constant (but rapidly sign changing) acceleration.

%====================================================================
\section{Negative frequency content of an inertial plane wave}\label{sec:Unruh:Integral}
%====================================================================
To make connection with the standard approach to computing the Unruh temperature\cite{Birrell_Davies:1982,Kiefer:1999,Gelis:2021}, in this section we compute the negative frequency content of a purely positive inertial plane wave as measured by the arbitrary accelerated observer.
We consider the case of when the acceleration profile $a(t)$ is nearly Rindler-like constant acceleration, 
and as such, consider the profile $a(t) = a_0\,\tanh(a_0\,t)$ considered in the previous \Sec{sec:Example:2}.

%====================================================================
\subsection{Constant acceleration}\label{subsec:Constant:acceleration}
%====================================================================
The inertial Minkowski plane waves are of the form $\frac{1}{\sqrt{\om}}\,e^{\mp\,i\,\om\,U}$, where $U=T-X$ is a lightcone coordinate (with corresponding $V=T+X$), and the minus/plus sign is for pure negative/positive frequency waves, respectively. 
For the accelerated observer (located at their origin $x=0$ with metric 
$ds^2 = -dt^2 + dx^2$), the purely positive and negative frequency plane waves are given by 
$\frac{1}{\sqrt{\Om}}\,\left.e^{\mp\,i\,\Om\,u}\right|_{x=0} = \frac{1}{\sqrt{\Om}}\,e^{\mp\,i\,\Om\,t}$ (with $u=t-x$, $v=t+x$). 
The essence of the analysis leading to the Unruh temperature is that the accelerated observer Rob will observe negative frequency components
$e^{i\,\Om\,t}$ in inertial Bob's purely positive frequency plane wave $e^{-i\,\om\,U}$. 

Bob's pure positive frequency plane wave be be Fourier analyzed by Rob as 
\bea{alpha:beta:WKB}
\hspace{-0.65in}
\tfrac{1}{\sqrt{\om}}\,e^{-i \om U} = 
\tfrac{1}{\sqrt{\Om}}\,\int_{-\infty}^{\infty} d\Om\,
\left(
\alpha_{\om \Om}\,e^{-i \Om t} + \beta_{\om \Om}\,e^{i \Om t}
\right)
&\Rightarrow&
\begin{bmatrix}
\alpha_{\om \Om} \\
\beta^*_{\om \Om} \\
\end{bmatrix}
= \frac{1}{2 \pi}\,\sqrt{\frac{\Om}{\om}}\,
\int_{-\infty}^{\infty} dt\, e^{\mp i \om U}\, e^{i \Om t} \defn \mA_\mp, \no 
S_\mp&\defn&  
\begin{bmatrix}
|\alpha_{\om \Om}|^2 \\
|\beta_{\om \Om}|^2 \\
\end{bmatrix}
\propto 
|\mA_\mp|^2.
%
%\hspace{-0.65in}
%&\Rightarrow&
%\begin{bmatrix}
%|\alpha_{\om \Om}|^2 \\
%|\beta_{\om \Om}|^2 \\
%\end{bmatrix}
%\propto
%S_\mp\defn
%\left| 
%\int_{-\infty}^{\infty} dt\, \ e^{i \Om t\, \mp\, i\,\left(\tfrac{\om}{a_0}\right)\,e^{-a_0 t }} 
%\right|^2 \defn |\mA_\mp|^2,\qquad
\eea
where the Fourier transform has been performed with respect to Rob's proper time $t$.
$|\alpha_{\om \Om}|^2$ and $|\beta_{\om \Om}|^2$ give the spectrum $S_\mp$ of the positive and negative 
frequency components (upper/lower signs) of Bob's purely positive frequency plane wave, as perceived by Rob.
The crucial point here is that Rob takes the Fourier transform $\mA$ with respect to \tit{his} proper time $t$, which is different than Bob's proper time $T$, due to the former's acceleration. 

If Rob were to be moving at constant velocity $\b_0$ (zero acceleration), with respect to Bob, then the relationship between the $(T,X)$ and $(t,x)$ coordinates is just the usual constant velocity Lorentz transformation so that
$T = \g_0\,(t + \b_0\,x)$ and $X = \g_0\,(x + \b_0\,t)$ so that $U=T-X = \sqrt{\tfrac{1-\b_0}{1+\b_0}}\,(t-x)$. Evaluating at $x=0$, we have the amplitude $\mA_\mp = \delta\big(\Om \mp \sqrt{\tfrac{1-\b_0}{1+\b_0}}\,\om\big)$. That is, from measuring $\mA_-$ (with $\om, \Om>0$), Rob observes  the frequency $\om$ of Bob's pure positive frequency wave to be simply doppler shifted $\Om = \sqrt{\tfrac{1\mp\b_0}{1\pm\b_0}}\,\om$  if he is moving away/towards Bob (upper/lower signs). Rob measures no negative frequency components $\mA_+=0$, since the argument of the delta function is purely positive.
For anything other than zero acceleration, Rob will measure $\mA_+\ne0$ namely, 
$|\beta_{\om \Om}|^2$ will be the measure of the negative frequency content of Bob's purely positive frequency wave, as observed by Rob\cite{Birrell_Davies:1982}.

For the Rindler constant acceleration case $a(t)=a_0$ we have $U=T-X = -\frac{1}{a_0}\, e^{-a_0\,t}$ and hence
\be{A:star:plus}
\mA^*_\mp \propto \int_{-\infty}^{\infty} dt\, \ e^{i \Om t\, \pm\, i\,\left(\tfrac{\om}{a_0}\right)\,e^{-a_0 t }}. 
\ee
The  integral can be performed by the change of variable $y=e^{-a_0 t}$ so that
$t=-1/a_0 \ln y$, and we then write $e^{i \Om t} = y^{-i \Om/a_0}$. We can then use the Gamma function integral formula (see p50 of Keifer \cite{Kiefer:1999}, and integral (3.381.5), p318 of Gradshteyn and Ryzhik \cite{Gradshteyn_Ryzhik:1965}) 
\bea{Keifer:p50:WKB}
\hspace{-0.25in}
\int_{y=0}^{\infty} dy\, y^{\nu-1}\, e^{-(A + i\,B)\,y} = \frac{\G(\nu)}{(A+ i B)^{\nu}} =
 \frac{\G(\nu)}{(A^2 + B^2)^{\nu/2}} \, e^{-i\,\nu\,\tan^{-1}(B/A)}
\;\;\overset{A\to 0}{\to}\;\;  \G(\nu)\,|B|^{-\nu} \, e^{-i\,\nu\,\tfrac{\pi}{2}\,\trm{sign}(B)},\quad
\eea
where $\G(\nu) = \int_{0}^{\infty} dy\, y^{\nu-1}\, e^{-y}$ is the Gamma function.
One then has $\nu = -i\tfrac{\Om}{a_0}$, $A=0$ and $B = \mp \om/a_0$ for $\mA_\mp$, 
leading to
\bea{Planck:spectrum:WKB}
\mA_\mp &=& \G(\tfrac{-i\Om}{a_0}) e^{\mp\,\tfrac{\pi \Om}{2\,a_0} }\,e^{i \big(\tfrac{\Om}{a_0}\big) \ln(\om/a_0)},
\\
\Rightarrow |\beta_{\om \Om}|^2 &\propto& |\G(\tfrac{-i\Om}{a_0})|^2 e^{-\tfrac{\pi \Om}{\,a_0} } 
= \tfrac{2 \pi}{(\Om/a_0)}\, \frac{1}{e^{2 \pi \Om/a_0}-1}
\equiv  \tfrac{2 \pi}{(\Om/a_0)}\, \frac{1}{e^{\hbar \Om/(k_b T_U)}-1},  \label{Planck:spectrum:WKB:line2} \\
\trm{using}\quad |\G(\nu)|^2 &=& \frac{\pi}{\nu\,\sinh(\pi\,\nu)},
\quad\trm{and defining}\quad
k_b T_U \defn \frac{\hbar\,a_0}{2\,\pi\,c}. \label{Planck:spectrum:WKB:line3}
\eea
after re-inserting the dimensional constants $k_b$ and $c$ in the Unruh temperature $T_U$, which we see by 
\Eq{Planck:spectrum:WKB:line3} has the thermal Planck spectrum for bosons. The lesson here is that 
\Eq{Planck:spectrum:WKB:line3}   reveals that Rob does \tit{not} perceive Bob's purely positive frequency plane wave as a purely positive frequency Rindler plane wave, and the latter interprets the negative frequency component spectrum $|\beta_{\om \Om}|^2 \ne 0$ as a bath of thermal particles with temperature proportional to his acceleration $a_0$. 
In the gravitational case, Rob would undergo constant acceleration if he sits stationary, at fixed radius, outside a black hole of mass $M$ with  Schwarzschild radius $R_s = 2 G M/c^2$.
The relevant acceleration Rob feels is the surface gravity $\kappa = GM/R_s^2$, and so he perceives a
Hawking temperature given by 
$T_H = T_U(a_0\to\kappa) = \frac{\hbar\,\kappa}{2\,\pi\,c} = \frac{\hbar\,c^3}{8\,\pi\,G\,M}$.

%=========================================================================
\subsection{$\mathbf{a(t) = a_0\,\boldsymbol{\tanh}(a_0\,t)}$}\label{subsec:Example2:acceleration}
%=========================================================================
We now return to the acceleration profile $a(t) = a_0\,\tanh(a_0\,t)$ in \Sec{sec:Example:2}, where
$2\,a_0\,T(\eta) = \sinh(\eta) + 2\,(\tan^{-1}(e^\eta) - \pi/4)$, and 
$2\,a_0\,(X(\eta)-X_0) = \sinh(\eta) - 2\,(\tan^{-1}(e^\eta) - \pi/4)$,
where $\eta = a_0\,t$,
so that 
$U = T-X = (X_0 + \frac{\pi}{2\,a_0}) + \frac{2}{a_0}\,(\tan^{-1}(e^\eta) - \pi/4)$.
The first term in parentheses is independent of $\eta$, and will lead to an irrelevant pure phase factor that will square to unity when we consider $S_+ = |A^*_+|^2$, so we'll drop it.
Defining $z=e^{-\eta}$ and 
considering the expansion of $\tan^{-1}(\tfrac{1}{z})$ about $z=0$ ($\eta\to\infty$),  we find
$U = -\frac{2}{a_0}\,e^{-a_0\,t} + \frac{2}{a_0}\,\sum_{k=1}^\infty (-1)^{2 k}\, \tfrac{e^{-(2k+1)\,a_0\,t}}{2k+1} \defn U_1 + \sum_{k=1}\,U_{2 k+1}$.
The first term $U_1 = -\frac{2}{a_0}\,e^{-a_0\,t}$ is twice the Rindler term appearing in \Eq{A:star:plus}.
We will treat the rest of the terms in the summation as corrections and expand 
$e^{i\,\tilde{\om}\,\sum_{k=1}\,U_{2 k+1}} \approx 1 +  i\,\tilde{\om}\,\sum_{k=1}\,U_{2 k+1} + \ldots$,
where we have defined $\tilde{\om} = \frac{2\,\om}{a_0}$. 
Therefore, upon making the usual Rindler subsitution $y = e^{-a_0\,t}$, we have 
to lowest order 
\be{A:star:plus:Ex:2}
A^*_+ = \int_{y=0}^{\infty} dy\, y^{\nu_0-1}\, e^{i\,\tilde{\om}\,y} 
\Big[
1+ i\,\tilde{\om} \,\sum_{k=1}^\infty \frac{(-1)^{2 k}\, y^{2 k+1}}{2k+1}
\Big],
\ee
where $\nu_0 = \frac{\Om}{a_0}$.
The first term $1$ in the square brackets is the usual Rindler term, as in \Eq{Planck:spectrum:WKB}.
The rest of the correction terms can be computed using the Gamma-integral formula \Eq{Keifer:p50:WKB},
now with $\nu = \nu_0 + (2k+1)$, for each $k\in\{1,2,\ldots\}$. Using the property of the Gamma function
$\G(z+1) = z\,\G(z)$, we have iteratively that $\G(\nu_0+2k+1) = (\nu_0)_{2k+1}\,\G(\nu_0)$, where
$(\nu_0)_{2k+1}$ is the Pochhammer symbol defined previously at the end of \Sec{sec:Example:2}.
Thus, for all the correction terms, we can factor out the term $\G(\nu_0) = \G(-i\frac{\Om}{a_0})$
that appears in \Eq{Planck:spectrum:WKB} which characterizes the constant acceleration Rindler term amplitude.
In fact, at any order in the expansion in \Eq{A:star:plus:Ex:2} the integrals will involve some power of
$y^{(\nu_0+p)-1}$ resulting in a contribution $\G(\nu_0+p) = (\nu_0)_{p}\,\G(\nu_0)$, so that $\G(\nu_0)$ can be
factored out to \tit{all} orders. The lesson of this calculation is that the spectrum 
$|\beta_{\om \Om}|^2 \propto |\mA_+|^2\sim |\G(\tfrac{-i\Om}{a_0})|^2 e^{-\tfrac{\pi \Om}{\,a_0} } $ 
of negative frequencies will contain the same Planck factor 
$\frac{1}{e^{2 \pi \Om/a_0}-1} \equiv \frac{1}{e^{\hbar \Om/(k_b T_U)}-1}$
as in \Eq{Planck:spectrum:WKB:line2} multiplied by unity plus a series of corrections depending on powers of $\tilde{\om}$ and $(\nu_0)_{p}$, i.e. a modified thermal spectrum. Note that the change in frequency
$\tilde{\om} = 2\om/a_0$ does not affect the Unruh temperature, which arise from the  phase factor 
$e^{i\,\Om\,t}$ containing $\Om$ (the frequency Rob detects).

%=========================================================================
\subsection{Arbitrary accleration}\label{subsec:arbitrary:acceleration}
%=========================================================================
One can an expansion similar to \Eq{A:star:plus:Ex:2} for a general acceleration $a(t)$.
Noting from \Eq{T:X:at:xeq0} (at $x=0$) that $U = (T-X)|_{x=0} = \int\, dt\,e^{-\chi(t)}$ one can write
$a(t) = a_0 + (a(t)-a_0)\defn a_0 + \d a(t)$, and expand in powers of $\d a(t)$ considered as small.
Then $e^{-\chi(t)} = e^{-a_0\,t} \,e^{-\D(t)} = e^{-a_0\,t} \,\sum_{n=0}^\infty (-1)^n\,\D(t)^n/n!$ 
where $\D(t) = \int^t\, dt'\,\d a(t')$. We then have
\bea{A:star:plus:general}
\mA^*_\mp &=& \int_{-\infty}^{\infty} dt\, e^{i (\Om\,t \mp \om\,U)}
=  \int_{-\infty}^{\infty} dt\, e^{i (\Om\,t \mp \om\,\int^t\, dt'\,\chi(t'))}, \no
&=& 
 \int_{-\infty}^{\infty} dt\, e^{i \big(\Om\,t \pm \om\,\frac{1}{a_0}\, e^{-a_0\,t}\big)}
 \, e^{\mp\,\om\, \sum_{n=1}^\infty\, F_n(t)}, \quad 
 F_n(t)\defn \int^t dt' 
 %\tfrac{(-i)^n\,e^{-a_0\,t}\,\D^n(t')}{n!}
\tfrac{(-i)^n}{n!}\, e^{-a_0\,t}\,\D^n(t').
\eea
Again, the term in the large parentheses in \Eq{A:star:plus:general} represents the constant acceleration Rindler integral leading to 
the usual Planck factor for the spectrum of negative frequencies $|\mA^*_+|^2$.
To lowest order in $\D(t)$ we can approximate the second exponential in  \Eq{A:star:plus:general}
as $1 \mp \om\, F_1(t)$ where $F_1(t) = i \int^t dt' \,e^{-a_0\,t'} \int^{t''} \d a(t'')$. Applying this to our first  example in \Sec{sec:Example:1} with acceleration profile $a(t) = a_0\, \cosh^{-2}(a_0\,t)$ produces an expansion qualitatively similar to that of
\Eq{A:star:plus:Ex:2}, except with an additional term in the expansion proportional to $y\ln y$ where
$y = \frac{2}{\pi}\chi(t)$. This term can be argued as small at the $y$ lower limit of integration $y=0$, and also small at the upper integration limit since the term $e^{\mp i\,y}$ (coming from the substitution $y=e^{-a_0\,t}$ varies infinitely rapidly at large values of $y$, and can safely be ignored (to lowest order). The remaining terms are polynomial in $y$ as in \Eq{A:star:plus:Ex:2} (and discussion thereafter), so that again one can factor out the term $\G(\nu_0) = \G(-i \tfrac{\Om}{a_0})$ responsible for producing the Unruh Planck factor $\frac{1}{e^{2 \pi \Om/a_0}-1} \equiv \frac{1}{e^{\hbar \Om/(k_b T_U)}-1}$ for the case constant acceleration.

Finally, note that a similar expansion of $\mF = \,a(t)\,e^{i\,y(t)} \int^t\, dt'\, e^{-i\,y(t')}$ 
%defined in \Eq{TU:at:line2}, 
that appears in the
denominator of the generalized Unruh temperature in \Eq{TU:at:line2},
can also be carried out as an expansion in $\d a(t)$ about the constant acceleration $a_0$.	
%====================================================================
\section{Conformal coordinates for arbitrary acceleration}\label{sec:Conformal:Coords}
%====================================================================
It is well known that any $1+1$ metric $ds^2 = -f^2(t,x)\,dt^2 + g^2(t,x)\,dx^2$
is locally flat\cite{Gomes:2025} and
can be transformed into a conformally flat metric of the form
$ds^2 = \Om_c^2\,(-d\tau^2+d\zeta^2)$, where $\Om_c^2$ is the conformal factor\cite{Birrell_Davies:1982}.
The importance of the conformal form of the metric is that one can readily infer that the plane waves of the associated Klein-Gordon wave equation are of the (inertial) form
$\frac{1}{\sqrt{\om}}\,e^{\mp i\,(\om\,\tau \mp k\,\zeta)}$ with $\om = |k|$, 
where $e^{\mp i\,\om\,\tau}$ corresponds to positive/negative frequencies (upper/lower signs) with respect to the observer's proper time $\tau$, 
and $e^{\pm i\,k\,\zeta}$ corresponds to right/left moving  waves (upper/lower signs), 
analogous to the inertial plane waves $\frac{1}{\sqrt{\om_i}}\,e^{\pm i\,(\om_i\,T \mp k_i\,X)}$ for the metric 
$ds^2 = -dT^2 + dX^2$. In this section we show how one can explicitly obtain the conformal form of the metric for arbitrary acceleration, and obtain explicit formulas for the transformation between the inertial $(T,X)$ and accelerated $(\tau, \zeta)$ coordinates.

We begin, in general, by noting that in $1+1$ dimensions we can trivially write
\be{metric:factorization}
ds^2 = -f^2(t,x)\,dt^2 + g^2(t,x)\,dx^2 \equiv -\big(f(t,x)\,dt - g(t,x)\,dx\big)\,\big(f(t,x)\,dt + g(t,x)\,dx\big). \no
\ee
We then define coordinates $(u,v)$ viz
\bea{u:v}
du &\defn& \mu_1(t,x)\,\big(f(t,x)\,dt - g(t,x)\,dx\big) = \frac{\pd u}{\pd t}\,dt + \frac{\pd u}{\pd x}\,dx, \label{u:v:line1} \\
dv &\defn& \mu_2(t,x)\,\big(f(t,x)\,dt + g(t,x)\,dx\big) = \frac{\pd v}{\pd t}\,dt + \frac{\pd v}{\pd x}\,dx, \label{u:v:line2}
\eea
where $\mu_1(t,x)$ and  $\mu_2(t,x)$  are integrating factors, necessary to ensure that
$du$ and $dv$ are total differentials. With the above definitions, the metric becomes
\bea{conformal: metric}
ds^2 &=& -\frac{1}{\mu_1\, \mu_2}\,du\,dv \equiv  \Om_c^2\,(-d\tau^2 + d\zeta^2), \\
\Om_c^2 &=& \frac{1}{\mu_1\, \mu_2}, \quad 
u= \tau-\zeta,\;  v= \tau+\zeta \Leftrightarrow 
\tau = \half(v+u),\;  \tau = \half(v-u),
\eea
where we have introduce temporal $\tau$  and spatial $\zeta$ coordinates 
via the lightcone coordinates $u$ and $v$.

 For $du$ and $dv$ to be total differentials,
  integrability conditions must be imposed that require  the cross partials to be equal,
namely that $\pd_x( \pd_t u) = \pd_t( \pd_x u)$ and $\pd_x( \pd_t v) = \pd_t( \pd_x v)$.
From \Eq{u:v:line1} and \Eq{u:v:line2}  with 
$(\pd_t u, \pd_x u) = (\mu_1\, f, -\mu_1\, g)$ and 
$(\pd_t v, \pd_x u) = (\mu_2\, f, \mu_2\, g)$
 this yields
\be{cross:partials}
\frac{\pd  }{\pd x}(\mu_1\,f) = -\frac{\pd }{\pd t}(\mu_1\,g ), \qquad  
\frac{\pd  }{\pd x}(\mu_2\,f) = \frac{\pd }{\pd t}(\mu_2\,g ).
\ee
Given $f$ and $g$, these are two separate equations for the unknown functions $\mu_1$ and $\mu_2$.
Once these solutions are know, one can write the solution to a general differential expression $du$ 
in two equivalent forms given a starting point $(t_0,x_0)$ as
\bea{u:t:x}
du \defn F(t,x)\,dt + G(t,x)\,dx \Rightarrow 
u(t,x) &=& \int_{t_0}^{t} dt' \, F(t',x) + \int_{x_0}^{x} dx'  \,G(t_0,x'), \label{u:t:x:line1} \\
         &=& \int_{t_0}^{t} dt'  \,F(t',x_0) + \int_{x_0}^{x} dx'  \,G(t,x').\label{u:t:x:line2} 
\eea
The above form in \Eq{u:t:x:line1} follows from noting that $\pd_t u(t,x) = F(t,x)$ as required.
Then $\pd_x u = \int_{t_0}^{t} dt' \, \pd_x F(t',x) + G(t_0,x) = 
 \int_{t_0}^{t} dt' \, \pd_t G(t',x) + G(t_0,x) = G(t,x)-G(t_0,x)+G(t_0,x) = G(t,x)$ as required,
 where we have used the integrability condition $\pd_x F(t',x) = \pd_t G(t',x)$ in the second equality, within the integral. The form of $u(t,x)$ in \Eq{u:t:x:line2} follows similarly, and the same form holds for integrating $dv$ to obtain $v(t,x)$.
 
 Turning to our metric for an accelerated observer $ds^2 = -(1 + a(t)\,x)^2\,dt^2 + dx^2$, we
 have $f(t,x) = 1 + a(t)\,x$ and $g(t,x)=1$. \Eq{cross:partials} becomes
 \bea{mu:solns}
  \hspace{-0.5in}
\frac{\pd  }{\pd x}\big(\mu_1\,(1 + a(t)\,x)\big) &=& -\frac{\pd }{\pd t}\,\mu_1 \;\;\Rightarrow\;\; 
\mu_1(t,x) = e^{-\chi(t)} \,\left(e^{-\chi(t)}\, x - \int^t dt' \, e^{-\chi(t')}\right) \equiv \mu_1(\chi,x), \quad \label{mu:solns:line1} \\
 \hspace{-0.5in}
\frac{\pd  }{\pd x}\big(\mu_2\,(1 + a(t)\,x)\big) &=& \;\;\; \frac{\pd }{\pd t}\,\mu_2 \;\;\Rightarrow\;\; 
\mu_2(t,x) = e^{\chi(t)} \,\left(e^{\chi(t)}\, x + \int^t dt' \, e^{\chi(t')}\right) \equiv \mu_2(\chi,x). \quad\label{mu:solns:line2} 
 \eea
 Invoking \Eq{u:t:x:line1}, and the definitions of $\tau$ and $\zeta$ we have
% \bea{u:v:final}
% \hspace{-0.5in}
% \tau(t,x) &=& 
% \half\,\left[
% \int_{t_0}^{t} dt'\,\big(\mu_2(t',x) + \mu_1(t',x)\big)\, (1+ a(t')\,x)  
% +  \int_{x_0}^{x} dx'\, \big(\mu_2(t_0,x') - \mu_1(t_0,x')\big)
% \right],
% \eea
%
 \bea{u:v:final}
 \hspace{-0.5in}
 \tau(t,x) &=& 
 \int_{t_0}^{t} dt'\,\left(\frac{\mu_2(t',x) + \mu_1(t',x)}{2}\right)\, \big(1+ a(t')\,x\big)  
 +  \int_{x_0}^{x} dx'\, \left(\frac{\mu_2(t_0,x') - \mu_1(t_0,x')}{2}\right), \\
  \hspace{-0.5in}
 \zeta(t,x) &=& 
 \int_{t_0}^{t} dt'\,\left(\frac{\mu_2(t',x) - \mu_1(t',x)}{2}\right)\, \big(1+ a(t')\,x\big)  
 +  \int_{x_0}^{x} dx'\, \left(\frac{\mu_2(t_0,x') + \mu_1(t_0,x')}{2}\right). 
 \eea
 
 For the case of the constant acceleration Rindler metric $ds_R^2 = -(1+a_0\,x)^2\,dt^2 + dx^2$ we find 
 using the above formulas
 \bea{Rindler:conformal:results}
\hspace{-0.5in}
 \mu_1 &=& e^{-2\,a_0\,t}\,(1+a_0\,x), \quad 
 \mu_2 = e^{2\,a_0\,t}\,(1+a_0\,x),\quad 
 \Om_c^2 = \frac{1}{\mu_1\,\mu_2} = \frac{1}{(1+a_0\,x)^2}, \label{Rindler:conformal:results:line1} \\
 \hspace{-0.5in}
 u &=& \frac{1- e^{-2\,a_0\,t}\,(1+a_0\,x)^2}{2\,a_0}, \quad 
 v = \frac{-1+ e^{2\,a_0\,t}\,(1+a_0\,x)^2}{2\,a_0}, \label{Rindler:conformal:results:line2} \\
 \hspace{-0.5in}
 \tau &=& \frac{v+u}{2} = \frac{\sinh(2\, a_0\,t)\,(1+a_0\,x)^2}{2\,a_0} \quad
 \zeta = \frac{v+u}{2} = \frac{-1+\cosh(2\,a_0\,t)\,(1+a_0\,x)^2}{2\,a_0}, \label{Rindler:conformal:results:line3} \\
 \hspace{-0.5in}
 ds^2 &=& \Om_c^2\, (-d\tau^2 + d\zeta^2) = 
 \frac{1}{(1+a_0\,x)^2} \, \left[(1+a_0\,x)^2\, \Big(-(1+a_0\,x)^2\,dt^2 + dx^2\Big)  \right] = -(1+a_0\,x)^2\,dt^2 + dx^2.\qquad \label{Rindler:conformal:results:line4} 
 \eea 
 
 Note that this is not the same as discussed in \Sec{sec:Rindler metric} where one typically defines the new spatial coordinate
 $\zeta'$ by $dx = (1+a_0\,x)\,d\zeta'$
or $(1+a_0\,x) = e^{a_0\,\zeta'}$, and the temporal coordinate $\tau'=t$,
which allows us to put the Rindler transformation into the form
of \Eq{Rindler:coords:t:x}, with the conformally flat form
%\be{Rindler:metric:conformally:flat}
$ds_R^2 = e^{2 a_0 \zeta'}\,(-d\tau'^2 + d\zeta'^2)$
with $\Om_c^{'2} =  e^{2 a_0 \zeta'}$.
Here $\tau'=t$ is purely a function of $t$, and $\zeta'$ is purely a function of $x$, while in 
\Eq{Rindler:conformal:results:line3}, $\tau$ and $\zeta$ are functions of both $t$ and $x$.
The method outlined above is valid  in general when the metric coefficients depend on \tit{both} the temporal and spatial coordinates.

Finally, recall that the coordinates $(t,x)$ of the arbitrarily accelerated observer are related in general to the inertial observer with coordinates $(T,X)$ and flat metric $ds^2 = -dT^2 + dX^2$
by \Eq{T:X:at:x:line1} and \Eq{T:X:at:x:line2} For the constant acceleration Rindler case with $T=\frac{1}{a_0}\,\sinh(a_0\,t)$ and 
$X=\frac{1}{a_0}\,\cosh(a_0\,t)$, 
and this choice of conformal coordinates $(\tau, \zeta)$ in \Eq{Rindler:conformal:results:line3},
straightforward algebraic manipulation and hyperbolic double angle identities yield
the correspondence $a_0\,\tau = (a_0\,X)\,(a_0\,T)$, and $1+2\,a_0\,\zeta = (a_0\,X)^2 + (a_0\,T)^2$, with
the relation $(1+2\,a_0\,\zeta)^2 - 2\,(a_0\,\tau)^2 = (a_0\,X)^4+ (a_0\,T)^4$.

%====================================================================
\section{Conclusion and Future Work}\label{sec:Conclusion}
%====================================================================
In this work we have generalized the standard form of the constant acceleration $a_0$ Rindler metric in flat Minkowski spacetime to arbitrary acceleration $a(t)$, which has the simple form of $a_0\to a(t)$ in the former. We developed the coordinate transformation between the inertial observer with coordinates $(T, X)$ and the accelerated observer with coordinates $(t, x)$, given an arbitrary acceleration profile $a(t)$ (or conversely, an arbitrary velocity profile $\b(t)$ respecting special relativity $|\b(t)| \le 1$). What emerges is the prominence of the integrated acceleration $\chi = \int dt\, a(t)$. We generalized the Unruh temperature derived from a WKB approach which treats the Unruh radiation (originally derived for constant acceleration) as arising from a quantum mechanical tunneling process, where the contribution to the temperature comes from integrating around the pole of the action, which occurs at the particle horizon. The generalized Unruh temperatures as the same form as the case of constant acceleration except now with (i) $a_0\to a(t)$, and (ii) a correction denominator that depends on and integrated exponential of $\chi$. We examined this generalized Unruh temperature for two acceleration profiles that had a short duration region of acceleration, and another with a hyperbolic tangent profile, so that it was essential constant at large times, but transitioned suddenly from negative to positive constant acceleration.
We made connection with the standard approach to calculating the Unruh temperature by means of examining the accelerated observer's measured negative frequency content of a purely positive frequency inertial plane wave. Lastly, we developed the explicit  transformation between the accelerated observer's coordinates and those for a  conformally flat metric, the latter of which allows us to most easily see the plane wave structure of the solutions of the wave equation for the accelerated observer as analogous in form to the of the inertial observer.

An area for future research includes the study of the response of detectors on arbitrary accelerated orbits in the spirit of Birrell and Davies classic reference \tit{Quantum Fields in Curved Space}\cite{Birrell_Davies:1982}.
In this regard the recent work of Hari, Barman and Kothawala\cite{Hari_Barman_Dawood:2025} is of note, which examines a piecewise accelerated trajectory, including regions of both positive and negative constant acceleration (of which our second examined example is an analogous continuous version). The author's also examine entanglement between the twin quantum detectors, showing that the changes in the direction of the acceleration leave an imprint on the detector responses and entanglement. It would be interesting to examine the effect of continuous versions of such arbitrary acceleration studied in this work on the entanglement (although the integrals would be much more difficult to mange, than those arising from the those author's piecewise profile).

Another area of interest for this work is  its extension to non-stationary metrics in curved spacetime, and an analogous formula for the Hawking temperature. Hari and Kothawala (2021) recent work\cite{Hari_Dawood:2021} examined the effect of tidal curvature on the dynamics of accelerated probes by developing (to lowest order) a coordinate transformation between an inertial observer on a geodesic path, and that of an accelerated observer, by means of an covariant power series expansion in Riemann normal coordinates. The authors develop a modified Unruh temperature in which 
$a_0\to \sqrt{a^2_0\,- \mathcal{E}_n}$, where $\mathcal{E}_n = R_{abcd}\,\eps^{ab}\,\eps^{cd}$ is a purely tidal component of curvature with $\eps^{ab}$ the bi-normal to the plane of motion. Interestingly, the
formula they develop for the transformation between the proper time of the inertial, geodesic observer and the accelerated observer is \tit{exactly} the form of the  formula developed 
by this author in a recent 2026 paper\cite{Alsing_I2R:2026} between a flat spacetime inertial observer and an accelerated observer with velocity profile $\b(t) = \b_0\,\tanh(a_0\,t) = \tanh(\chi(t))$ with $|\b_0|<1$. Most likely, this correspondence arises due to the instantaneous Lorentz transformation from the inertial geodesic observer and the accelerated observer using the instantaneous velocity profile $\b(t)$, though role of tidal term  $ \mathcal{E}_n$, and the reason it acts as a negative acceleration to $a_0$ would be of interest for further investigation.

Finally, it would be of particular interest to examine the case in curved spacetime of an observer at fixed radial coordinate remaining stationary outside a collapsing collapsing shell of matter (in the process of forming a black hole). Such an observer would no longer experience constant acceleration to remain stationary as the shell collapsed, but would undergo a time dependent acceleration, depending on the details of the collapse. 
Again, the role of both the acceleration profile $a(t)$, and it's possible modifications by tidal effects would be of keen interest to study.

%================================
%\clearpage
%\newpage
%================================
\begin{acknowledgments}
The author would like to thank K. Hari and  D. Kothawala for informative discussions on their work.

The author has no competing interests for this work. \newline
\indent The author intends to make data openly available in the near future at \newline 
%{\verb+<to be filled in if published>+}, 
{\verb+https://dataverse.harvard.edu/dataverse/alsingpm_research+},
including the 
\tit{Mathematica} codes that were used to construct 
\Fig{fig:a0divcoshSqrda0t}  and \Fig{fig:a0tanha0t}.
%} % end red
%{\verb+https://dataverse.harvard.edu/dataverse/alsingpm_research+}. 
\end{acknowledgments}
%================================

%================================
%% Create the reference section using BibTeX:
%%\nocite{apsrev41Control}
%%\bibliographystyle{ieeetr}
%%\bibliographystyle{apsrev4-1}
%\bibliography{alsing_composite_bibfile}

\begin{thebibliography}{31}%
\makeatletter
\providecommand \@ifxundefined [1]{%
 \@ifx{#1\undefined}
}%
\providecommand \@ifnum [1]{%
 \ifnum #1\expandafter \@firstoftwo
 \else \expandafter \@secondoftwo
 \fi
}%
\providecommand \@ifx [1]{%
 \ifx #1\expandafter \@firstoftwo
 \else \expandafter \@secondoftwo
 \fi
}%
\providecommand \natexlab [1]{#1}%
\providecommand \enquote  [1]{``#1''}%
\providecommand \bibnamefont  [1]{#1}%
\providecommand \bibfnamefont [1]{#1}%
\providecommand \citenamefont [1]{#1}%
\providecommand \href@noop [0]{\@secondoftwo}%
\providecommand \href [0]{\begingroup \@sanitize@url \@href}%
\providecommand \@href[1]{\@@startlink{#1}\@@href}%
\providecommand \@@href[1]{\endgroup#1\@@endlink}%
\providecommand \@sanitize@url [0]{\catcode `\\12\catcode `\$12\catcode
  `\&12\catcode `\#12\catcode `\^12\catcode `\_12\catcode `\%12\relax}%
\providecommand \@@startlink[1]{}%
\providecommand \@@endlink[0]{}%
\providecommand \url  [0]{\begingroup\@sanitize@url \@url }%
\providecommand \@url [1]{\endgroup\@href {#1}{\urlprefix }}%
\providecommand \urlprefix  [0]{URL }%
\providecommand \Eprint [0]{\href }%
\providecommand \doibase [0]{http://dx.doi.org/}%
\providecommand \selectlanguage [0]{\@gobble}%
\providecommand \bibinfo  [0]{\@secondoftwo}%
\providecommand \bibfield  [0]{\@secondoftwo}%
\providecommand \translation [1]{[#1]}%
\providecommand \BibitemOpen [0]{}%
\providecommand \bibitemStop [0]{}%
\providecommand \bibitemNoStop [0]{.\EOS\space}%
\providecommand \EOS [0]{\spacefactor3000\relax}%
\providecommand \BibitemShut  [1]{\csname bibitem#1\endcsname}%
\let\auto@bib@innerbib\@empty
%</preamble>
\bibitem [{\citenamefont {Unruh}(1976)}]{Unruh:1976}%
  \BibitemOpen
  \bibfield  {author} {\bibinfo {author} {\bibfnamefont {W.~G.}\ \bibnamefont
  {Unruh}},\ }\href@noop {} {\bibfield  {journal} {\bibinfo  {journal} {Phys.
  Rev. D}\ }\textbf {\bibinfo {volume} {14}},\ \bibinfo {pages} {870} (\bibinfo
  {year} {1976})}\BibitemShut {NoStop}%
\bibitem [{\citenamefont {de~Gill}\ \emph {et~al.}(2010)\citenamefont
  {de~Gill}, \citenamefont {Singleton}, \citenamefont {Akhmedova},\ and\
  \citenamefont {Pilling}}]{deGill:2010}%
  \BibitemOpen
  \bibfield  {author} {\bibinfo {author} {\bibfnamefont {A.}~\bibnamefont
  {de~Gill}}, \bibinfo {author} {\bibfnamefont {D.}~\bibnamefont {Singleton}},
  \bibinfo {author} {\bibfnamefont {V.}~\bibnamefont {Akhmedova}}, \ and\
  \bibinfo {author} {\bibfnamefont {T.}~\bibnamefont {Pilling}},\ }\href@noop
  {} {\bibfield  {journal} {\bibinfo  {journal} {Am. J. Phys.}\ }\textbf
  {\bibinfo {volume} {78}},\ \bibinfo {pages} {685} (\bibinfo {year}
  {2010})}\BibitemShut {NoStop}%
\bibitem [{\citenamefont {Hawking}(1975)}]{Hawking:1975}%
  \BibitemOpen
  \bibfield  {author} {\bibinfo {author} {\bibfnamefont {S.~W.}\ \bibnamefont
  {Hawking}},\ }\href@noop {} {\bibfield  {journal} {\bibinfo  {journal}
  {Commun. Math. Phys.}\ }\textbf {\bibinfo {volume} {43}},\ \bibinfo {pages}
  {199} (\bibinfo {year} {1975})}\BibitemShut {NoStop}%
\bibitem [{\citenamefont {Birrell}\ and\ \citenamefont
  {Davies}(1982)}]{Birrell_Davies:1982}%
  \BibitemOpen
  \bibfield  {author} {\bibinfo {author} {\bibfnamefont {N.~D.}\ \bibnamefont
  {Birrell}}\ and\ \bibinfo {author} {\bibfnamefont {P.~C.~W.}\ \bibnamefont
  {Davies}},\ }\href@noop {} {\emph {\bibinfo {title} {Quantum fields in curved
  space}}}\ (\bibinfo  {publisher} {Cambridge University Press, Cambridge},\
  \bibinfo {year} {1982})\BibitemShut {NoStop}%
\bibitem [{\citenamefont {Unruh}\ and\ \citenamefont
  {Wald}(1982)}]{Unruh_Wald:1982}%
  \BibitemOpen
  \bibfield  {author} {\bibinfo {author} {\bibfnamefont {W.~G.}\ \bibnamefont
  {Unruh}}\ and\ \bibinfo {author} {\bibfnamefont {R.~M.}\ \bibnamefont
  {Wald}},\ }\href@noop {} {\bibfield  {journal} {\bibinfo  {journal} {Phys.
  Rev. D}\ }\textbf {\bibinfo {volume} {25}},\ \bibinfo {pages} {942} (\bibinfo
  {year} {1982})}\BibitemShut {NoStop}%
\bibitem [{\citenamefont {Crispino}\ \emph {et~al.}(2008)\citenamefont
  {Crispino}, \citenamefont {Higuchi},\ and\ \citenamefont
  {Matas}}]{Crispino:2008}%
  \BibitemOpen
  \bibfield  {author} {\bibinfo {author} {\bibfnamefont {L.~C.}\ \bibnamefont
  {Crispino}}, \bibinfo {author} {\bibfnamefont {A.}~\bibnamefont {Higuchi}}, \
  and\ \bibinfo {author} {\bibfnamefont {G.~E.~A.}\ \bibnamefont {Matas}},\
  }\href@noop {} {\bibfield  {journal} {\bibinfo  {journal} {Rev. Mod. Phys.}\
  }\textbf {\bibinfo {volume} {80}},\ \bibinfo {pages} {787} (\bibinfo {year}
  {2008})}\BibitemShut {NoStop}%
\bibitem [{\citenamefont {Alsing}\ and\ \citenamefont
  {Milonni}(2004)}]{Alsing_Milonni:2004}%
  \BibitemOpen
  \bibfield  {author} {\bibinfo {author} {\bibfnamefont {P.~M.}\ \bibnamefont
  {Alsing}}\ and\ \bibinfo {author} {\bibfnamefont {P.~W.}\ \bibnamefont
  {Milonni}},\ }\href@noop {} {\bibfield  {journal} {\bibinfo  {journal} {Am.
  J. Phys.}\ }\textbf {\bibinfo {volume} {72}},\ \bibinfo {pages} {1524}
  (\bibinfo {year} {2004})}\BibitemShut {NoStop}%
\bibitem [{\citenamefont {Padmanabhan}(2010)}]{Padmanabhan:2010}%
  \BibitemOpen
  \bibfield  {author} {\bibinfo {author} {\bibfnamefont {T.}~\bibnamefont
  {Padmanabhan}},\ }\href@noop {} {\emph {\bibinfo {title} {Gravitation:
  Foundations and Frontiers}}}\ (\bibinfo  {publisher} {Cambridge University
  Press, Cambridge},\ \bibinfo {year} {2010})\BibitemShut {NoStop}%
\bibitem [{\citenamefont {Gelis}(2021)}]{Gelis:2021}%
  \BibitemOpen
  \bibfield  {author} {\bibinfo {author} {\bibfnamefont {F.}~\bibnamefont
  {Gelis}},\ }\href@noop {} {\emph {\bibinfo {title} {Problems in Quantum Field
  Theory with fully-worked solutions}}}\ (\bibinfo  {publisher} {Cambridge
  University Press, Cambridge},\ \bibinfo {year} {2021})\ pp.\ \bibinfo {pages}
  {339--347}\BibitemShut {NoStop}%
\bibitem [{\citenamefont {Scully}\ and\ \citenamefont
  {Zubairy}(1997)}]{Scully_Zubairy:1997}%
  \BibitemOpen
  \bibfield  {author} {\bibinfo {author} {\bibfnamefont {M.~O.}\ \bibnamefont
  {Scully}}\ and\ \bibinfo {author} {\bibfnamefont {M.~S.}\ \bibnamefont
  {Zubairy}},\ }\href@noop {} {\emph {\bibinfo {title} {Quantum Optics, (Chap.
  9)}}}\ (\bibinfo  {publisher} {Cambridge University Press},\ \bibinfo
  {address} {Cambridge},\ \bibinfo {year} {1997})\BibitemShut {NoStop}%
\bibitem [{\citenamefont {Agarwal}(2013)}]{Agarwal:2013}%
  \BibitemOpen
  \bibfield  {author} {\bibinfo {author} {\bibfnamefont {G.~S.}\ \bibnamefont
  {Agarwal}},\ }\href@noop {} {\emph {\bibinfo {title} {Quantum Optics}}}\
  (\bibinfo  {publisher} {Cambridge University Press},\ \bibinfo {address}
  {Cambridge},\ \bibinfo {year} {2013})\BibitemShut {NoStop}%
\bibitem [{\citenamefont {Gerry}\ and\ \citenamefont
  {Knight}(2023)}]{Gerry_Knight:2023}%
  \BibitemOpen
  \bibfield  {author} {\bibinfo {author} {\bibfnamefont {C.~C.}\ \bibnamefont
  {Gerry}}\ and\ \bibinfo {author} {\bibfnamefont {P.~L.}\ \bibnamefont
  {Knight}},\ }\href@noop {} {\emph {\bibinfo {title} {Introductory Quantum
  Optics, 2nd Ed.}}}\ (\bibinfo  {publisher} {Cambridge University Press,
  Cambridge},\ \bibinfo {year} {2023})\BibitemShut {NoStop}%
\bibitem [{\citenamefont {Carroll}(2004)}]{Carroll:2004}%
  \BibitemOpen
  \bibfield  {author} {\bibinfo {author} {\bibfnamefont {S.~M.}\ \bibnamefont
  {Carroll}},\ }\href@noop {} {\emph {\bibinfo {title} {Spacetime and
  Geometry}}}\ (\bibinfo  {publisher} {Addison Wesley, San Francisco},\
  \bibinfo {year} {2004})\BibitemShut {NoStop}%
\bibitem [{\citenamefont {Srinivasan}\ and\ \citenamefont
  {Padmanabhan}(1999)}]{Srinivasan:1999}%
  \BibitemOpen
  \bibfield  {author} {\bibinfo {author} {\bibfnamefont {K.}~\bibnamefont
  {Srinivasan}}\ and\ \bibinfo {author} {\bibfnamefont {T.}~\bibnamefont
  {Padmanabhan}},\ }\href@noop {} {\bibfield  {journal} {\bibinfo  {journal}
  {Phys. Rev. D}\ }\textbf {\bibinfo {volume} {60}},\ \bibinfo {pages} {024007}
  (\bibinfo {year} {1999})}\BibitemShut {NoStop}%
\bibitem [{\citenamefont {Shankaranarayanan}\ \emph {et~al.}(2001)\citenamefont
  {Shankaranarayanan}, \citenamefont {Padmanabhan},\ and\ \citenamefont
  {Srinivasan}}]{Srinivasan:2001}%
  \BibitemOpen
  \bibfield  {author} {\bibinfo {author} {\bibfnamefont {S.}~\bibnamefont
  {Shankaranarayanan}}, \bibinfo {author} {\bibfnamefont {T.}~\bibnamefont
  {Padmanabhan}}, \ and\ \bibinfo {author} {\bibfnamefont {K.}~\bibnamefont
  {Srinivasan}},\ }\href@noop {} {\bibfield  {journal} {\bibinfo  {journal}
  {Class. Quant. Gravity}\ }\textbf {\bibinfo {volume} {19}},\ \bibinfo {pages}
  {2671} (\bibinfo {year} {2001})},\ \Eprint {http://arxiv.org/abs/1407.1704}
  {1407.1704} \BibitemShut {NoStop}%
\bibitem [{\citenamefont {Parikh}\ and\ \citenamefont
  {Wilczek}(2000)}]{Wilczek:2000}%
  \BibitemOpen
  \bibfield  {author} {\bibinfo {author} {\bibfnamefont {M.~K.}\ \bibnamefont
  {Parikh}}\ and\ \bibinfo {author} {\bibfnamefont {F.}~\bibnamefont
  {Wilczek}},\ }\href@noop {} {\bibfield  {journal} {\bibinfo  {journal} {Phys.
  Rev. Lett.}\ }\textbf {\bibinfo {volume} {85}},\ \bibinfo {pages} {5042}
  (\bibinfo {year} {2000})}\BibitemShut {NoStop}%
\bibitem [{\citenamefont {Vanzo}\ \emph {et~al.}(2011)\citenamefont {Vanzo},
  \citenamefont {Acquaviva},\ and\ \citenamefont {Criscienzo}}]{Vanzo:2011}%
  \BibitemOpen
  \bibfield  {author} {\bibinfo {author} {\bibfnamefont {L.}~\bibnamefont
  {Vanzo}}, \bibinfo {author} {\bibfnamefont {G.}~\bibnamefont {Acquaviva}}, \
  and\ \bibinfo {author} {\bibfnamefont {R.~D.}\ \bibnamefont {Criscienzo}},\
  }\href@noop {} {\bibfield  {journal} {\bibinfo  {journal} {Classical and
  Quantum Gravity}\ }\textbf {\bibinfo {volume} {28}},\ \bibinfo {pages}
  {183001} (\bibinfo {year} {2011})}\BibitemShut {NoStop}%
\bibitem [{\citenamefont {Zhang}\ \emph {et~al.}(2025)\citenamefont {Zhang},
  \citenamefont {Corda},\ and\ \citenamefont {Cai}}]{Corda:2025}%
  \BibitemOpen
  \bibfield  {author} {\bibinfo {author} {\bibfnamefont {B.}~\bibnamefont
  {Zhang}}, \bibinfo {author} {\bibfnamefont {C.}~\bibnamefont {Corda}}, \ and\
  \bibinfo {author} {\bibfnamefont {Q.}~\bibnamefont {Cai}},\ }\href@noop {}
  {\bibfield  {journal} {\bibinfo  {journal} {Entropy}\ }\textbf {\bibinfo
  {volume} {27}},\ \bibinfo {pages} {167} (\bibinfo {year} {2025})}\BibitemShut
  {NoStop}%
\bibitem [{\citenamefont {Kerner}\ and\ \citenamefont
  {Mann}(2008)}]{Kerner_Mann:2008}%
  \BibitemOpen
  \bibfield  {author} {\bibinfo {author} {\bibfnamefont {R.}~\bibnamefont
  {Kerner}}\ and\ \bibinfo {author} {\bibfnamefont {R.~B.}\ \bibnamefont
  {Mann}},\ }\href@noop {} {\bibfield  {journal} {\bibinfo  {journal}
  {Classical and Quantum Gravity}\ }\textbf {\bibinfo {volume} {25}},\ \bibinfo
  {pages} {095014} (\bibinfo {year} {2008})}\BibitemShut {NoStop}%
\bibitem [{\citenamefont {Mann}(2025)}]{Mann:2015}%
  \BibitemOpen
  \bibfield  {author} {\bibinfo {author} {\bibfnamefont {R.~B.}\ \bibnamefont
  {Mann}},\ }\href@noop {} {\emph {\bibinfo {title} {Black Holes:
  Thermodynamics, Information and Firewalls}}}\ (\bibinfo  {publisher}
  {Springer International Publishing},\ \bibinfo {year} {2025})\BibitemShut
  {NoStop}%
\bibitem [{\citenamefont {Hari}\ and\ \citenamefont
  {Kothawala}(2021)}]{Hari_Dawood:2021}%
  \BibitemOpen
  \bibfield  {author} {\bibinfo {author} {\bibfnamefont {K.}~\bibnamefont
  {Hari}}\ and\ \bibinfo {author} {\bibfnamefont {D.}~\bibnamefont
  {Kothawala}},\ }\href@noop {} {\bibfield  {journal} {\bibinfo  {journal}
  {Phys. Rev. D}\ }\textbf {\bibinfo {volume} {104}},\ \bibinfo {pages}
  {064032} (\bibinfo {year} {2021})}\BibitemShut {NoStop}%
\bibitem [{\citenamefont {Hari}\ \emph {et~al.}()\citenamefont {Hari},
  \citenamefont {Barman},\ and\ \citenamefont
  {Kothawala}}]{Hari_Barman_Dawood:2025}%
  \BibitemOpen
  \bibfield  {author} {\bibinfo {author} {\bibfnamefont {K.}~\bibnamefont
  {Hari}}, \bibinfo {author} {\bibfnamefont {S.}~\bibnamefont {Barman}}, \ and\
  \bibinfo {author} {\bibfnamefont {D.}~\bibnamefont {Kothawala}},\ }\href@noop
  {} {\bibinfo  {journal} {arxiv:2512.10908, [gr-qc]}\ }\BibitemShut {NoStop}%
\bibitem [{\citenamefont {Rindler}(1964)}]{Rindler:1956}%
  \BibitemOpen
\bibfield  {journal} {  }\bibfield  {author} {\bibinfo {author} {\bibfnamefont
  {W.}~\bibnamefont {Rindler}},\ }\href@noop {} {\bibfield  {journal} {\bibinfo
   {journal} {Am. J. Phys}\ }\textbf {\bibinfo {volume} {34}},\ \bibinfo
  {pages} {1174} (\bibinfo {year} {1964})}\BibitemShut {NoStop}%
\bibitem [{\citenamefont {Alsing}(2026)}]{Alsing_I2R:2026}%
  \BibitemOpen
  \bibfield  {author} {\bibinfo {author} {\bibfnamefont {P.~M.}\ \bibnamefont
  {Alsing}},\ }\href@noop {} {\bibfield  {journal} {\bibinfo  {journal}
  {arxiv:2601.18956}\ } (\bibinfo {year} {2026})}\BibitemShut {NoStop}%
\bibitem [{\citenamefont {D'Inverno}(1992)}]{DInverno:1992}%
  \BibitemOpen
  \bibfield  {author} {\bibinfo {author} {\bibfnamefont {R.}~\bibnamefont
  {D'Inverno}},\ }\href@noop {} {\emph {\bibinfo {title} {Introducing
  Einstein's Relativity}}}\ (\bibinfo  {publisher} {Oxford University Press,
  London UK},\ \bibinfo {year} {1992})\BibitemShut {NoStop}%
\bibitem [{\citenamefont {Ryder}(2009)}]{Ryder:2009}%
  \BibitemOpen
  \bibfield  {author} {\bibinfo {author} {\bibfnamefont {L.}~\bibnamefont
  {Ryder}},\ }\href@noop {} {\emph {\bibinfo {title} {Introduction to General
  Relativity}}}\ (\bibinfo  {publisher} {Cambridge University Press, Cambridge,
  UK},\ \bibinfo {year} {2009})\BibitemShut {NoStop}%
\bibitem [{\citenamefont {Landau}\ and\ \citenamefont
  {Lifshitz}(1975)}]{Landau_Lifshitz:CTF:1975}%
  \BibitemOpen
  \bibfield  {author} {\bibinfo {author} {\bibfnamefont {L.~D.}\ \bibnamefont
  {Landau}}\ and\ \bibinfo {author} {\bibfnamefont {E.~M.}\ \bibnamefont
  {Lifshitz}},\ }\href@noop {} {\emph {\bibinfo {title} {The Classical Theory
  of Fields, 4th Ed.}}}\ (\bibinfo  {publisher} {Pergamon Press Ltd., Elmsford,
  NY},\ \bibinfo {year} {1975})\ pp.\ \bibinfo {pages} {234--237}\BibitemShut
  {NoStop}%
\bibitem [{\citenamefont {Alsing}(1998)}]{Alsing_AJP:1998}%
  \BibitemOpen
  \bibfield  {author} {\bibinfo {author} {\bibfnamefont {P.~M.}\ \bibnamefont
  {Alsing}},\ }\href@noop {} {\bibfield  {journal} {\bibinfo  {journal} {Am. J.
  Phys.}\ }\textbf {\bibinfo {volume} {66}},\ \bibinfo {pages} {779} (\bibinfo
  {year} {1998})}\BibitemShut {NoStop}%
\bibitem [{\citenamefont {Kiefer}(1999)}]{Kiefer:1999}%
  \BibitemOpen
  \bibfield  {author} {\bibinfo {author} {\bibfnamefont {C.}~\bibnamefont
  {Kiefer}},\ }in\ \href@noop {} {\emph {\bibinfo {booktitle} {Classical and
  Quantum Black Holes}}},\ \bibinfo {editor} {edited by\ \bibinfo {editor}
  {\bibfnamefont {P.}~\bibnamefont {Fr\'{e}}}, \bibinfo {editor} {\bibfnamefont
  {V.}~\bibnamefont {Gorini}}, \bibinfo {editor} {\bibfnamefont
  {G.}~\bibnamefont {Magli}}, \ and\ \bibinfo {editor} {\bibfnamefont
  {U.}~\bibnamefont {Moschella}}}\ (\bibinfo  {publisher} {IoP Press},\
  \bibinfo {address} {Philadelphia},\ \bibinfo {year} {1999})\ pp.\ \bibinfo
  {pages} {19--74}\BibitemShut {NoStop}%
\bibitem [{\citenamefont {Gradshteyn}\ and\ \citenamefont
  {Ryzhik}(1965)}]{Gradshteyn_Ryzhik:1965}%
  \BibitemOpen
  \bibfield  {author} {\bibinfo {author} {\bibfnamefont {I.~S.}\ \bibnamefont
  {Gradshteyn}}\ and\ \bibinfo {author} {\bibfnamefont {I.~M.}\ \bibnamefont
  {Ryzhik}},\ }\href@noop {} {\emph {\bibinfo {title} {{Table of Integrals,
  Series adn Products, 4th Ed.}}}}\ (\bibinfo  {publisher} {Academic Press, New
  York},\ \bibinfo {year} {1965})\BibitemShut {NoStop}%
\bibitem [{\citenamefont {Gomes}(2025)}]{Gomes:2025}%
  \BibitemOpen
  \bibfield  {author} {\bibinfo {author} {\bibfnamefont {H.}~\bibnamefont
  {Gomes}},\ }\href {\doibase doi:10.31389/pop.201} {\bibfield  {journal}
  {\bibinfo  {journal} {Philosophy of Physics}\ }\textbf {\bibinfo {volume}
  {3}},\ \bibinfo {pages} {17} (\bibinfo {year} {2025})}\BibitemShut {NoStop}%
\end{thebibliography}
%================================
% paste in the .bbl file here
\providecommand{\noopsort}[1]{}\providecommand{\singleletter}[1]{#1}%
%
%================================

%================================
\end{document}